\documentclass[a4paper,11pt]{article}

\usepackage{bbm}
\usepackage{longtable}
\usepackage{pdflscape}

\usepackage{amsfonts}
\usepackage{amssymb}
\usepackage{amsmath}
\usepackage{dsfont}
\usepackage[width=16.5cm]{geometry}
\usepackage{color}
\usepackage{ulem}
\usepackage{url}

\newcommand{\I}{\mathrm{i}}
\newcommand{\Id}{\ensuremath{1\!\!1}}
\newcommand{\Z}[1]{\ensuremath{\mathds{Z}_{#1}}} % z_N ->\Z{N}
\newcommand{\ZZ}{\ensuremath{\mathds{Z}}}  % integers
\newcommand{\QQ}{\ensuremath{\mathds{Q}}}  % rationals
\newcommand{\RR}{\ensuremath{\mathds{R}}}
\newcommand{\GL}[1]{\ensuremath{\mathrm{GL}(#1)}}  % general linear group
\newcommand{\SL}[1]{\ensuremath{\mathrm{SL}(#1)}}
\newcommand{\PSL}[1]{\ensuremath{\mathrm{PSL}(#1)}}

\newcommand{\Carat}{{\sc carat}}
\newcommand{\bmat}{B}                      % lattice basis matrix
\newcommand{\basis}[1]{\mathfrak{#1}}      % basis (latt. or euc.)

\newcommand{\U}[1]{\ensuremath{\mathrm{U}(#1)}}
\newcommand{\SU}[1]{\ensuremath{\mathrm{SU}(#1)}}
\newcommand{\SO}[1]{\ensuremath{\mathrm{SO}(#1)}}
\newcommand{\E}[1]{\ensuremath{\mathrm{E}_{#1}}} % e.g. \E{8}

\newcommand{\GAP}[1]{$\left[#1\right]$}
\newcommand{\rep}[1]{\ensuremath\boldsymbol{#1}}
\newcommand{\crep}[1]{\ensuremath\overline{\boldsymbol{#1}}}

\newcommand{\Apref}[1]{Appendix~\ref{#1}}
\newcommand{\Secref}[1]{Section~\ref{#1}}
\newcommand{\Eqref}[1]{Equation~\eqref{#1}}
\newcommand{\Eqsref}[2]{Equations~\eqref{#1} and~\eqref{#2}}

\newcommand{\Tabref}[1]{Table~\ref{#1}}

\definecolor{purple}{rgb}{0.5,0,0.5}
\begin{document}

\begin{titlepage}

\vspace*{-3.0cm}
\begin{flushright}
\normalsize{DESY-13-075}\\
\normalsize{TUM-HEP 887/13}\\
\normalsize{FLAVOUR(267104)-ERC-42}\\
\end{flushright}

\vspace*{1.0cm}

\begin{center}
{\Large\textbf{Heterotic non--Abelian orbifolds}}

\vspace{1cm}

\textbf{Maximilian~Fischer} \\[5mm]
\textit{\small~Physik--Department T30, Technische Universit\"at M\"unchen, James--Franck--Stra\ss e, 85748 Garching, Germany} \\[8mm]

\textbf{Sa\'ul~Ramos-S\'anchez} \\[5mm]
\textit{\small~Department of Theoretical Physics, Physics Institute, UNAM, Mexico D.F. 04510, Mexico} \\[8mm]

\textbf{Patrick~K.S.~Vaudrevange} \\[5mm]
\textit{\small~Deutsches Elektronen--Synchrotron DESY, Notkestra\ss e 85, 22607 Hamburg, Germany}
\end{center}

\vspace{1cm}

\vspace*{1.0cm}

\begin{abstract}
We perform the first systematic analysis of particle spectra obtained from 
heterotic string compactifications on non--Abelian toroidal orbifolds. After 
developing a new technique to compute the particle spectrum in the case of 
standard embedding based on higher dimensional supersymmetry, we compute the 
Hodge numbers for all recently classified 331 non--Abelian orbifold geometries 
which yield $\mathcal{N}=1$ supersymmetry for heterotic compactifications. 
Surprisingly, most Hodge numbers follow the empiric pattern 
$h^{(1,1)} - h^{(2,1)} = 0\text{ mod } 6$, which might be related to the number 
of three standard model generations.
Furthermore, we study the fundamental groups in order to identify the 
possibilities for non--local gauge symmetry breaking. Three examples are 
discussed in detail: the simplest non--Abelian orbifold $S_3$ and two more 
elaborate examples, $T_7$ and $\Delta(27)$, which have only one untwisted 
K\"ahler and no untwisted complex structure modulus. Such models might 
be especially interesting in the context of no--scale supergravity. Finally, we briefly 
discuss the case of orbifolds with vanishing Euler numbers in the context of 
enhanced (spontaneously broken) supersymmetry.
\end{abstract}
\end{titlepage}

\newpage

% ==============================================================================
\section{Introduction}

Ten--dimensional superstring theory is perhaps the most promising candidate
to yield an ultraviolet completion of particle physics and to explain some 
persisting cosmological puzzles. One useful mechanism to overcome the challenge 
of reducing the number of dimensions while preserving $\mathcal N=1$ 
supersymmetry and other phenomenologically appealing features is to compactify 
the six extra spatial dimensions on a toroidal orbifold.

Toroidal orbifolds offer a fairly simple geometrical structure that allows one 
to deal with the compactification in terms of the conformal--field framework of 
string theory~\cite{Dixon:1985jw,Dixon:1986jc}. Moreover, in recent years these 
constructions have become a fruitful source of semi--realistic models in the 
heterotic strings\footnote{References for other successful constructions are for 
example (see also references therein): for the free fermionic 
construction~\cite{Faraggi:1991jr}, for Gepner constructions~\cite{Dijkstra:2004cc}, 
for type II with D--branes~\cite{Gmeiner:2005vz,Blumenhagen:2006ci,Gmeiner:2008xq}, for 
M--theory on $G_2$ manifolds~\cite{Acharya:2008zi} and for Calabi--Yau 
spaces~\cite{Bouchard:2005ag,Anderson:2011ns}. For the connection between 
(singular) orbifolds and (smooth) Calabi--Yau compactifications, see 
e.g.~\cite{Blaszczyk:2010db,Blaszczyk:2011ig,Blaszczyk:2011hs,Buchmuller:2012mu}.}. The 
resulting scenarios can simultaneously reproduce the matter spectrum 
of the minimal supersymmetric version of the standard 
model~\cite{Buchmuller:2005jr,Buchmuller:2006ik,Lebedev:2006kn,Lebedev:2008un} 
(or its singlet extensions~\cite{Lebedev:2009ag}) and provide new approaches
for solving puzzles such as the existence of hierarchies~\cite{Antoniadis:1994hg,Kappl:2008ie}, 
family symmetries~\cite{Kobayashi:2006wq} and proton stability~\cite{Lee:2010gv,Forste:2010pf}.

Despite these encouraging features, not all possible toroidal orbifold 
geometries have been explored.
In the past, several efforts have led to partial classifications of orbifold geometries.
The first attempts were restricted to \Z{N} orbifolds~\cite{Kobayashi:1991rp,Bailin:1999nk}.
Much later, $\Z{2}\times\Z{2}$ orbifolds including so--called 
roto--translations\footnote{They are also known as {\it shift orbifolds}; see e.g. 
Ref.~\cite{Blumenhagen:2006ab} for type IIA string theory on shift 
$\Z{2}\times\Z{2}$ orientifolds.}
were successfully classified~\cite{Donagi:2008xy} (see also~\cite{Donagi:2004ht,Forste:2006wq}). 
However, it is evident that more general orbifolds, which include not only Abelian
(i.e. $\Z{N}$ and $\Z{M}\times\Z{N}$) but also non--Abelian point groups 
(for example, $S_n$ for $n=3,4$ and $D_{2n}$ for $n=2,3$), could also 
lead to appealing physics. Only recently a full classification of all 
(symmetric\footnote{For a recent work on asymmetric orbifolds, see 
e.g.~\cite{Beye:2013moa}.}) toroidal orbifold geometries which preserve 
$\mathcal{N}=1$ SUSY in the context of heterotic compactifications has been 
achieved~\cite{Fischer:2012qj}.

Even though most of the $\mathcal N=1$ heterotic orbifolds of 
Ref.~\cite{Fischer:2012qj} (331 out of 469) are based on the discrete action of
non--Abelian point groups, the geometrical aspects and phenomenology of these 
non--Abelian orbifolds have been studied only in few 
cases~\cite{Kakushadze:1996hj,Konopka:2012gy}. The purpose of this paper is to 
provide the first tools to address these questions.

With this goal in mind, after a brief description of orbifold 
compactifications of the heterotic string, we develop a technique to 
systematically determine all twisted sectors and their fixed points/tori of all
toroidal orbifolds in \Secref{sec:nonabelianorbifolds}. Further, for the 
so--called {\it standard embedding} of the orbifold action into the gauge 
degrees of freedom, which leads to an $\E6$ gauge group, we develop a technique 
based on supersymmetry in four and six dimensions to compute the number 
of $\rep{27}$ and $\crep{27}$ matter representations. This provides, as 
is known, the Hodge numbers $h^{1,1}$ and $h^{2,1}$, 
respectively. Furthermore, we determine the fundamental groups of all 
non--Abelian orbifolds in order to identify the possibility of non--local GUT 
breaking~\cite{Ross:2004mi,Hebecker:2004ce,Anandakrishnan:2012ii}.
\Secref{sec:examples} is devoted to a detailed study of 
three sample non--Abelian orbifolds, with point groups $S_3$, $T_7$ and $\Delta(27)$, 
which illustrate the main properties of these constructions.
Finally, we discuss our findings in \Secref{sec:Discussion}. Our most 
important results are summarized in \Tabref{tab:FundamentalGroups} for a 
list of all non--trivial fundamental groups and \Tabref{tab:NonAbelianPointGroups} 
for a list of Hodge numbers and their geometrical origin.

% ==============================================================================
\section{Heterotic non--Abelian orbifolds}
\label{sec:nonabelianorbifolds}
We consider compactifications of the ten--dimensional heterotic string on 
(symmetric) toroidal orbifolds~\cite{Dixon:1985jw,Dixon:1986jc}, where points on the 
six--dimensional torus $T^6$ are identified under the action of the so--called 
orbifolding group $G$, i.e.
\begin{equation}
\mathds{O} ~=~ T^6/G ~=~ \RR^6/S\;.
\end{equation}
Equivalently, the orbifold $\mathds{O}$ is defined as $\RR^6$ with an 
identification of points under the action of the so--called space group $S$. 
An element of $S$ consists of a rotational part and a translation. In detail,
\begin{equation}
g=(\vartheta, \lambda) \in S \quad\text{acts on $x\in\RR^6$ as}\quad g\; x = \vartheta\; x + \lambda\;,
\end{equation}
where $\lambda$ is a shift of $x$ in $\RR^6$ and $\vartheta\in\SU{3}\subset\SO{6}$ a rotation. 
Then, $g\; x \sim x$ for all $g \in S$ is the equivalence relation that defines 
$\mathds{O}$. The pure--translational elements of $S$ have the form $(\Id,\lambda)$,
where $\lambda$ can be expanded in terms of six basis vectors $e_i$ as 
$\lambda = n_i e_i$ with integer coefficients $n_i$ and summation over 
$i=1,\ldots,6$. Hence, the pure translations define a lattice 
$\Lambda$ and hereby the torus $T^6=\RR^6/\Lambda$. 
On the other hand, for $\vartheta\neq\Id$ there can additionally be elements 
$(\vartheta, \lambda) \in S$ with $\lambda\not\in\Lambda$ (i.e. with fractional $n_i$), 
which are called roto--translations~\cite{Hebecker:2004ce}.

The rotational part $\vartheta$ of all elements $g = (\vartheta,\lambda) \in S$ 
forms a group, the so--called point group $P$. We further define the orbifolding 
group $G$ as the group generated by all $g = (\vartheta,\lambda) \in S$,
where two elements that are related by a pure lattice translation are identified. 
Therefore, the orbifolding group $G$ is equivalent to the point group $P$ if 
no roto--translations are present.

The Abelian case is well studied, resulting in many phenomenological 
interesting models in, for example,
$\Z{6}$-II~\cite{Kobayashi:2004ya,Buchmuller:2005jr,Buchmuller:2006ik,Lebedev:2006kn,Lebedev:2008un}, 
$\Z{12}$-I~\cite{Kim:2007mt}, $\Z{2}\times\Z{2}$~\cite{Blaszczyk:2009in} and 
$\Z{2}\times\Z{4}$~\cite{Pena:2012ki} (there were also earlier phenomenological 
studies in \Z3 and \Z7, e.g.~\cite{Ibanez:1987pj,Casas:1990yt}). In this 
paper we deal with non--Abelian orbifolds, i.e. with orbifolds whose 
point groups $P$ are non--Abelian. We consider all inequivalent point groups 
(also known as {\it $\QQ$--classes}), all inequivalent lattices (called {\it $\ZZ$--classes}) 
and all roto--translations (i.e. \textit{affine classes})\footnote{For further details on
the definitions of these classes, we suggest Refs.~\cite{Brown:1978,Plesken:1998,Fischer:2012qj}.}. 
We use these six--dimensional spaces to compactify the 10D 
$\E{8}\times\E{8}'$ heterotic string to four dimensions. 

On heterotic orbifolds, there are two kinds of closed strings which contribute to the 
massless particle spectrum of the resulting four--dimensional effective theory: 
(i) untwisted strings that close already in flat $\RR^6$, and (ii) twisted strings that 
close only on the orbifold due to a non--trivial rotation $\vartheta$ (and
possibly a translation $\lambda$) in the respective boundary condition, 
e.g. a twisted string generated by $g=(\vartheta,\lambda)$ closes under the
boundary condition $X(\tau,\sigma+2\pi)=\vartheta\; X(\tau,\sigma) + 2\pi \lambda$
for the bosonic string coordinate. It follows that 
twisted strings are localized at the fixed points/fixed tori of the orbifold 
geometry. In the case of standard embedding (which is a specific choice of how the 
orbifold acts in the gauge degrees of freedom of the heterotic string), 
the matter spectrum consists of $\rep{27}$--, $\crep{27}$--plets and singlets
of the four--dimensional observable gauge group $\E{6}$. As we describe in more 
detail in the next section, counting the numbers of the non--trivial representations of
\E6 allows us to compute the Hodge numbers of these heterotic compactifications,
which is the primary purpose of this paper.

\subsection{Hodge numbers}
\label{sec:hodgeNumbers}

The Hodge numbers $(h^{(1,1)}, h^{(2,1)})$ count the K\"ahler and 
complex structure moduli, respectively, which correspond to deformations (of 
size and shape) of the geometry. For the heterotic orbifolds under 
consideration, we can split these numbers into contributions from the untwisted 
sector and from the twisted sectors,
\begin{equation}
(h^{(1,1)}, h^{(2,1)}) = (h_\text{U}^{(1,1)}, h_\text{U}^{(2,1)}) 
                       + (h_\text{T}^{(1,1)}, h_\text{T}^{(2,1)})\,,
\end{equation}
and compute them as we explain in the following.

\subsubsection{Contributions from the untwisted sectors}
In this section, we demonstrate that the number of untwisted moduli can be 
computed directly from the point group $P$ using representation 
theory of finite groups. 

The 4D untwisted K\"ahler and complex structure moduli, counted respectively by
$h_\text{U}^{(1,1)}$ and $h_\text{U}^{(2,1)}$, originate from the nine plus nine 
internal components of the 10D supergravity multiplet of the heterotic string,
which correspond to the following string excitations
\begin{subequations}
\label{eqn:untwistedmoduli}
\begin{eqnarray}
|q\rangle_\text{R} \otimes \tilde\alpha_{-1}^{\bar{\jmath}} |0\rangle_\text{L}  &&\qquad\text{for K\"ahler moduli,} \label{eqn:UntwistedKahlerModulus} \\
|q\rangle_\text{R} \otimes \tilde\alpha_{-1}^j |0\rangle_\text{L}          &&\qquad\text{for complex structure moduli,} \label{eqn:UntwistedComplexStructureModulus}
\end{eqnarray}
\end{subequations}
where $j=1,2,3$ and $|q\rangle_\text{R}$ denotes the ground state of the 
(supersymmetric) right--mover with (bosonized) momenta
\begin{equation}
q=(0,\underline{-1,0,0})\;,
\end{equation}
where the underline denotes permutations. Furthermore, $\tilde\alpha_{-1}^{\bar{\jmath}}$ or 
$\tilde\alpha_{-1}^j$ excites the left--moving ground state $|0\rangle_\text{L}$ 
in the $j$--th complex plane spanned by the complex coordinate $\bar{z}^{\bar{\jmath}}$ or $z^j$. 

On the orbifold only the invariant combinations of these (untwisted) states 
survive as unfixed moduli. As untwisted moduli are uncharged with respect to the 
gauge group, they transform only under the action of the point group $P$. From 
Table C.2 in Ref.~\cite{Fischer:2012qj} we know the explicit form of the point group as a 
three--dimensional, in general reducible representation $\rep{\rho}$ of $P$, 
with $P$ being a finite sub-group of $\SU{3}$. Under the action of the point 
group $P$ the right--moving ground state and the oscillator excitations 
transform as
\begin{subequations}
\label{eqn:trafountwistedmoduli}
\begin{eqnarray}
|q\rangle_\text{R}               & \text{ transforms as } & \rep{\rho} \\
\tilde\alpha_{-1}^{\bar{\jmath}} & \text{ transforms as } & \crep{\rho} \\
\tilde\alpha_{-1}^j              & \text{ transforms as } & \rep{\rho}\;.
\end{eqnarray}
\end{subequations}
Hence, using \Eqsref{eqn:untwistedmoduli}{eqn:trafountwistedmoduli}, one can 
count the number of untwisted moduli $(h_\text{U}^{(1,1)}, h_\text{U}^{(2,1)})$ 
from the tensor products
\begin{equation}
\label{eq:TensorProduct}
\rep{\rho} \otimes \crep{\rho} \;\rightarrow\; h_\text{U}^{(1,1)} \rep{\rho_0} \oplus \ldots \quad\text{and}\quad \rep{\rho} \otimes \rep{\rho} \;\rightarrow\; h_\text{U}^{(2,1)} \rep{\rho_0} \oplus \ldots \;,
\end{equation}
where $\rep{\rho_0}$ denotes the trivial singlet representation of $P$ and 
$h_\text{U}^{(1,1)}$ and $h_\text{U}^{(2,1)}$ are the multiplicities in the 
respective decomposition. These multiplicities can be computed most easily 
using characters (the character of an element $g \in P$ in the representation 
$\rep{\rho}$ is given by $\chi_{\rep{\rho}}(g) = \text{Tr}(\rep{\rho}(g))$). In 
general, a decomposition of a tensor product reads $\rep{\rho_1} \otimes \rep{\rho_2} 
= \bigoplus_{i=1}^c n_i \rep{\rho_i}$, where $c$ is the number of inequivalent 
irreducible representations, which are denoted as $\rep{\rho_i}$, and $n_i$ are 
the corresponding multiplicities. Then, 
$\chi_{\rep{\rho_1} \otimes \rep{\rho_2}}(g) = \sum_{i=1}^c n_i \chi_{\rep{\rho_i}}(g)$ 
and one can compute the multiplicities $n_i$ using the orthogonality of the rows 
of the character table, see e.g. Section 4.2 of Ref.~\cite{Fischer:2012qj}.

We use the software GAP~\cite{GAP4} and Mathematica to perform these computations. The results 
are listed in \Tabref{tab:SummaryUntwistedModuli}. Note that there are many
cases with only one untwisted K\"ahler modulus (i.e. only the overall 
volume of $\mathds{O}$ is unfixed) and no complex structure modulus (for 
example $P=T_7$, see \Secref{sec:T7Example}), which might be especially 
interesting in the context of no--scale 
supergravity~\cite{Ellis:1983sf,Cvetic:1988yw,Brignole:1997dp,Covi:2008ea}. This is in contrast to 
orbifolds with Abelian point groups where always at least the three K\"ahler 
moduli (associated with the sizes of the compact space split in three complex 
planes) survive the orbifold projection (see e.g.~\cite{Ibanez:1992hc}). 

\begin{table}[!ht]
\centering
\begin{tabular}{|c|l|}
\hline
untwisted moduli                           &  \\
$(h_\text{U}^{(1,1)}, h_\text{U}^{(2,1)})$ & non--Abelian point groups \\
\hline
\hline
(2,2) & $S_3$, $D_4$, $D_6$ \\
\hline
(2,1) & $QD_{16}$, $(\Z{4}\times\Z{2})\rtimes\Z{2}$, $\Z{4}\times S_3$, $\left(\Z{6}\times\Z{2}\right)\rtimes\Z{2}$, \\
      & $\GL{2,3}$, $\SL{2,3}\rtimes\Z{2}$ \\
\hline
(2,0) & $\Z{8}\rtimes\Z{2}$, $\Z{3}\times S_3$, $\Z{3}\rtimes\Z{8}$, $\SL{2,3}\mathrm{-I}$, $\Z{3}\times D_4$, \\
      & $\Z{3}\times Q_8$, $\left(\Z{4}\times\Z{4}\right)\rtimes\Z{2}$, $\Z{3}\times \left(\Z{3}\rtimes\Z{4}\right)$, $\Z{6}\times S_3$, \\
      & $\Z{3}\times\SL{2,3}$, $\Z{3}\times\left(\left(\Z{6}\times\Z{2}\right)\rtimes\Z{2}\right)$, $\SL{2,3}\rtimes\Z{4}$ \\
\hline
(1,1) & $A_4$, $S_4$ \\
\hline
(1,0) & $T_7$, $\Delta(27)$, $\Z{3}\times A_4$, $\Delta(48)$, $\Delta(54)$, $\Z{3}\times S_4$, $\Delta(96)$, \\
      & $\Sigma(36\phi)$, $\Delta(108)$, $\PSL{3,2}$, $\Sigma(72\phi)$, $\Delta(216)$ \\
\hline
\end{tabular}
\vspace{-1mm}
\caption{List of non--Abelian point groups with specified number of untwisted moduli 
$(h_\text{U}^{(1,1)}, h_\text{U}^{(2,1)})$. For example, orbifolds with point 
group $S_3$, $D_4$ and $D_6$ have two untwisted K\"ahler moduli and two 
untwisted complex structure moduli, i.e. $(h_\text{U}^{(1,1)}, h_\text{U}^{(2,1)})=(2,2)$.}
\label{tab:SummaryUntwistedModuli}
\end{table}

\subsubsection{Contributions from the twisted sectors}
The twisted sectors of the orbifold yield some twisted K\"ahler moduli (also known as
{\it blow--up modes}) and twisted complex structure moduli (which describe the shapes 
of unorbifolded fixed tori, as explained in more detail later).

In order to determine their numbers, we analyze the standard embedding of the 
$\E{8}\times\E{8}'$ heterotic string, which results in a four--dimensional 
$\mathcal{N}=1$ theory with $\E{6}\times\E{8}'$ gauge group~\cite{Konopka:2012gy} 
(for Abelian point groups, the gauge group includes additional model--dependent 
gauge factors, such as $\U{1}^2$, $\SU2\times\U1$ or \SU3). 
Due to the (2,2) world--sheet supersymmetry, the number of twisted 
$\rep{27}$--plets corresponds to $h_\text{T}^{(1,1)}$ and the number of twisted 
$\crep{27}$--plets gives $h_\text{T}^{(2,1)}$~\cite{Dine:1986zy,Dixon:1989fj}. 
In order to identify the number of twisted $\rep{27}$-- and $\crep{27}$--plets 
we first have to consider the orbifold fixed points and fixed tori in some 
detail, with a special focus on four-- and six--dimensional supersymmetry (see 
also Ref.~\cite{Nibbelink:2012de} for a related discussion).

\paragraph{Twisted sectors.}
In Abelian orbifolds, the twisted sectors are labeled by their point group 
elements, e.g. for a $\Z{M}\times\Z{N}$ point group with generators $\vartheta$ 
and $\omega$ we use $T_{k,\ell}$, with $k=0,\ldots,M-1$ and $\ell=0,\ldots,N-1$, 
to denote the (twisted) sector produced by $\vartheta^k \omega^\ell \in P$. 
In contrast, for non--Abelian orbifolds a twisted sector is characterized by a 
conjugation class $[\vartheta]$ for $\vartheta \in P$ and hence it is denoted 
as $T_{[\vartheta]}$.

\paragraph{Fixed points/tori.}
For a given twisted sector $T_{[\vartheta]}$, a space group element 
$g=(\vartheta,\lambda) \in S$ with $\vartheta \neq \Id$ is called a 
constructing element of a massless string if the fixed point equation 
associated with $g$,
\begin{equation}
g\; f = f \qquad\Leftrightarrow\qquad \vartheta\; f + \lambda = f \quad\text{for } f \in \RR^6\;,
\end{equation}
has a zero-- or a two--dimensional solution $f$. In the former case, 
$f$ is called a fixed point, while in the latter it is called a fixed torus.

\paragraph{Equivalence of fixed points/tori.}
Different solutions $f$ can be geometrically equivalent in the compact space
due to the symmetries induced by the compactification. Equivalences between 
the solutions are easily identified via their corresponding constructing elements.
We distinguish two different kinds of equivalences:
\begin{enumerate}
 \item[{\bf (i)}] {\bf Equivalence on the torus.}
Take two constructing elements of massless strings with the same point group 
element, i.e. $g_1=(\vartheta,\lambda_1) \in S$ and $g_2=(\vartheta,\lambda_2) 
\in S$. They are said to be equivalent on the torus if $g_1$ and $g_2$ are in 
the same conjugacy class with respect to translations, i.e. 
\begin{equation}
g_1 = h g_2 h^{-1} \quad\text{for some } h=(\Id,\lambda) \in S\;.
\end{equation}
In other words, $g_1\sim g_2$ on the torus if $\lambda\in\Lambda$ exists such 
that $\lambda_1-\lambda_2 = \left(\Id-\vartheta\right)\lambda$. Then, the 
corresponding fixed points/tori differ by a lattice vector. Using this 
definition of equivalence one can determine for each twisted sector 
$T_{[\vartheta]}$ all inequivalent constructing elements on the torus.

\item[{\bf (ii)}]{\bf Equivalence on the orbifold.}
Fixed points/tori that are inequivalent on the torus can be equivalent on 
the orbifold, i.e. fixed points/tori of a given twisted sector can be identified 
by a further orbifold action. Again, this equivalence can be determined using 
the concept of conjugacy classes, now allowing for general $h \in S$, i.e.
\begin{equation}
g_1 \sim g_2 \quad\text{if}\quad g_1,g_2 \in [g]=\{hgh^{-1} \text{ for all } h \in S\}\;.
\end{equation}
Fixed points/tori associated with elements of the same conjugacy class are 
identified on the orbifold. In more detail, take $g_1, g_2 \in [g]$ with
\begin{equation}
g_1\; f_1 = f_1 \quad\text{and}\quad g_2\; f_2 = f_2\;,
\end{equation}
where $f_1, f_2 \in \RR^6$ denote the fixed points/tori. As 
$g_1, g_2 \in [g]$ there exists an element $h \in S$ such that 
$g_2 = hg_1h^{-1}$. Then, $\left(h^{-1}f_2\right) = \left(h^{-1}g_2\right)f_2 = 
\left(g_1 h^{-1}\right)f_2 = g_1 \left(h^{-1}f_2\right)$. Consequently, 
$f_1 = h^{-1}f_2$ and we see that the fixed points/tori $f_1$ and $f_2$ are 
identified on the orbifold.
\end{enumerate}

\paragraph{Massless twisted matter.}
After obtaining all inequivalent constructing elements on the orbifold we start 
with the determination of the associated twisted matter spectrum. Each 
constructing element $g=(\vartheta, \lambda) \in S$ defines a boundary 
condition for a closed string on the orbifold, i.e.
\begin{equation}
Z(\tau, \sigma + 2\pi) = g\; Z(\tau, \sigma) = \vartheta\;Z(\tau, \sigma) + 2 \pi \lambda\;.
\end{equation}
For each twisted sector one can choose a basis of the three compactified complex 
coordinates $Z^i$ such that the twist $\vartheta \in \SU{3}$ becomes diagonal. 
Using this basis, the twist can be expressed by the so-called twist vector 
$v=(v_1,v_2,v_3)$, whose components are the rotational phases in units of $2\pi$. These twist 
vectors are analogous to the well--known ones for the case of Abelian point groups (see e.g. Table 1 in 
Ref.~\cite{Dixon:1986jc} and Table 5.2 in Ref.~\cite{Fischer:2012qj}). Furthermore, 
the gauge embedding can be diagonalized such that its action is parametrized by 
a shift, similarly as in the Abelian case. For the standard embedding, 
we choose $V=(v_1,v_2,v_3,0^5)(0^8)$. In order to compute the twisted matter,
one may need a basis change for each twisted sector of a non--Abelian orbifold. 
However, for each individual sector one can use the standard tools and 
the intuition developed from the well--known Abelian case, such as the
usual masslessness equations for left-- and right--movers and their solutions.

\paragraph{Invariance of twisted matter.}
In this way one can construct the Hilbert space $\mathcal H_{[g]}$ of massless 
twisted strings with constructing element $g$. However, not all states from 
$\mathcal H_{[g]}$ are necessarily invariant 
under the full orbifold action. One has to consider projections, i.e. one has 
to project the Hilbert space $\mathcal H_{[g]}$ of massless strings to the invariant 
subspace with respect to all space group elements $h$ that commute with the 
constructing element $g$, $gh=hg$. The set of commuting elements is called the 
centralizer of $g$. It is important to note that the rotational part of $g$ and 
$h$ and their gauge embeddings can be diagonalized simultaneously, as they 
commute.

For each constructing element $g \in [g]$ one distinguishes two cases:
\begin{enumerate}
\item In the first case, $g$ is related to a fixed point (not a fixed torus). 
Then, ten--dimensional $\mathcal{N}=1$ supersymmetry is broken down to 
$\mathcal{N}=1$ in four dimensions at the fixed point of $g$, and the 
Hilbert space $\mathcal H_{[g]}$ only respects 4D $\mathcal{N}=1$. 
Fixed points with these properties contribute only one twisted 
$\rep{27}$--plet, which can be related to one twisted K\"ahler modulus (i.e. blow--up 
mode), but no $\crep{27}$--plet and therefore no twisted complex structure 
modulus. In other words, the constructing element $g$ yields a contribution 
$(1,0)$ to the Hodge numbers $(h_\text{T}^{(1,1)}, h_\text{T}^{(2,1)})$. 
Let us point out that $\mathcal H_{[g]}$ and $\mathcal H_{[g^{-1}]}$ are not
independent, since $\mathcal H_{[g^{-1}]}$ contains the CPT conjugate partners 
of $\mathcal H_{[g]}$. Thus, it suffices to consider only $\mathcal H_{[g]}$ 
in the computations.

\item In the second case, $g$ has a fixed torus. Considering only the
action of $g$ (and $g^{-1}$) on the internal space, the theory on this
fixed torus has $\mathcal{N}=1$ in six dimensions (i.e. 4D
$\mathcal{N}=2$) with $\E{7}$ observable gauge group and a twisted
$\rep{56}$ hypermultiplet (or half--hypermultiplet). In terms of 4D
$\mathcal{N}=1$ this twisted $\rep{56}$--plet originates from the
sector $g$ contributing a left--chiral superfield, which transforms as
$\rep{56}$ of $\E{7}$, and from the sector $g^{-1}$ contributing
another left--chiral superfield, which transforms in the complex
conjugate representation, e.g. as a $\rep{56}$--plet with negative
$\U{1}$ charge. However, in the case $g=g^{-1}$ (or $[g]=[g^{-1}]$)
the twisted $\rep{56}$--plet is real, e.g. a $\rep{56}$--plet with
zero $\U{1}$ charge. Hence, it transforms as a half--hypermultiplet.

From the full 4D perspective the $\E{7}$ is broken to $\E{6}$ and the
left--chiral $\rep{56}$--plet from $g$ branches into $\rep{27} \oplus
\crep{27}$ plus two singlets. Thus, 4D matter originates from
$\mathcal{N}=2$ (half--)hypermultiplets and in terms of 4D
$\mathcal{N}=1$ a constructing element $g$ with fixed torus
contributes both, one twisted $\rep{27}$--plet and one twisted
$\crep{27}$--plet, to the Hilbert space $\mathcal H_{[g]}$. This would
result in one twisted K\"ahler modulus and one twisted complex
structure modulus per fixed torus. 

However, in the whole orbifold one has to perform the projection on invariant states: 
if there is (at least) one element in the centralizer of $g$ which breaks $\mathcal{N}=1$ 
in six dimensions to $\mathcal{N}=1$ in four dimensions, the twisted 
$\crep{27}$--plet is removed from $\mathcal H_{[g]}$ and, consequently, the twisted 
complex structure modulus of this orbifolded fixed torus is projected out. 
Then, the fixed torus of $g$ contributes $(1,0)$ to the Hodge numbers. On the 
other hand, if all 
elements of the centralizer keep $\mathcal{N}=1$ in six dimensions, the 
twisted $\rep{27}$-- and $\crep{27}$--plet and hence the respective moduli 
survive this projection. In this case, the fixed torus is not orbifolded 
further by the action of the centralizer and the twisted complex structure 
modulus describes the shape of this torus. Then, the fixed torus of $g$ 
contributes $(1,1)$ to the Hodge numbers.
\end{enumerate}

Based on these observations, we notice that it is enough to know the 
geometrical aspects (space group, constructing elements, etc.) of the orbifold 
and not the details of the gauge embedding in order to arrive at the
Hodge numbers. As a test, we have first used this procedure to corroborate the Hodge numbers for all 138 
orbifolds with Abelian point groups of Ref.~\cite{Fischer:2012qj} (originally 
obtained using the {\small\tt orbifolder}~\cite{Nilles:2011aj}). 
Then, we applied this procedure to the 331 orbifolds with non--Abelian 
point groups. The results are listed in \Tabref{tab:NonAbelianPointGroups} of
\Apref{sec:NonAbelianResults}. We discuss three examples in detail in 
\Secref{sec:examples}. It is interesting to note that, like
in the Abelian case of Ref.~\cite{Fischer:2012qj},
also the Hodge numbers of most non--Abelian cases 
satisfy the empiric rule
\begin{equation}
h^{(1,1)} - h^{(2,1)} = 0 \text{ mod } 6\;,
\label{eq:empiricRelation}
\end{equation}
for which we have not found an explanation yet (see also Ref.~\cite{Bailin:1999nk}). 
In the cases where~\Eqref{eq:empiricRelation}
is satisfied and the Euler number $\chi=2(h^{(1,1)} - h^{(2,1)})$ does not vanish, it seems
conceivable that the addition of discrete Wilson lines~\cite{Dixon:1986jc,Ibanez:1986tp}
can lead to candidate models with three generations of standard model
particles.

\subsection{Fundamental group}
The fundamental group $\pi_1$ of a toroidal orbifold is given by the following 
quotient group \cite{Dixon:1985jw,Brown:2002}
\begin{equation}
\pi_1 ~=~ S/\langle F\rangle\;,
\end{equation}
where $S$ is the space group that defines the orbifold, $F$ is the set of all 
constructing elements and $\langle F\rangle$ is the group generated by the 
elements of $F$.

There are two possible origins for a generator of $\pi_1$: either it arises 
from a roto--translation (i.e. from the orbifolding group $G$) or from a pure 
translation (i.e. from the lattice $\Lambda$). In order to identify this, we 
compute in addition to $\pi_1 = S/\langle F\rangle$ also 
$G/G_F$ and $\Lambda/\Lambda_F$, where $G_F \subset G$ is generated by the 
roto--translations of $\langle F\rangle$ and $\Lambda_F\subset\Lambda$ is the 
lattice of $\langle F\rangle$.

In total we find that 38 out of 331 orbifolds with non--Abelian point group 
and $\mathcal{N}=1$ have a non--trivial fundamental group, for example 
$\pi_1=\Z{2}$, $\Z{3}$, $\Z{2}\times\Z{2}$ and $\Z{3}\times\Z{3}$. They are listed in 
\Tabref{tab:FundamentalGroups} of \Apref{sec:NonAbelianResults}. In the 
next section we discuss one of them in detail. Combined with the results of 
\cite{Fischer:2012qj} we have a complete list of (toroidal, $\mathcal{N}=1$) 
orbifold geometries which offer a non--trivial fundamental group: there are 
69 cases out of 469. These cases are of special interest for 
phenomenology as they may allow for non--local GUT breaking 
\cite{Ross:2004mi,Hebecker:2004ce,Anandakrishnan:2012ii}. Therefore, the 
gauge embeddings of the (freely--acting) elements of the fundamental group 
and the conditions from modular invariance must be analyzed, 
cf.~\cite{Blaszczyk:2010db}.

\paragraph{Example: $\boldsymbol{D_4}$ orbifold.}
Let us discuss the case $D_4$--1--5 (i.e. \ZZ--class \#1, affine class \#5) 
with Hodge numbers $(6,6)$ in detail. $D_4$ is generated by $\vartheta$ and 
$\omega$ fulfilling $\vartheta^2=\omega^2=(\vartheta\omega)^4=\Id$. In the case 
(1--5) the space group $S$ is generated by 
$g_1=(\vartheta,\frac{1}{2} e_1+\frac{1}{4} e_5)$ and $g_2=(\omega, 0)$, where
\begin{equation}
\vartheta_\basis{e} ~=~ \left(
\begin{array}{cccccc}
 1 & 0 & 0 & 0 & 0 & 0 \\
 0 & 0 & 0 & -1 & 0 & 0 \\
 0 & 0 & -1 & 0 & 0 & 0 \\
 0 & -1 & 0 & 0 & 0 & 0 \\
 0 & 0 & 0 & 0 & -1 & 0 \\
 0 & 0 & 0 & 0 & 0 & -1 \\
\end{array}
\right) \quad\text{and}\quad
\omega_\basis{e} ~=~ \left(
\begin{array}{cccccc}
 0 & 0 & 1 & 0 & 0 & 0 \\
 0 & 1 & 0 & 0 & 0 & 0 \\
 1 & 0 & 0 & 0 & 0 & 0 \\
 0 & 0 & 0 & -1 & 0 & 0 \\
 0 & 0 & 0 & 0 & -1 & 0 \\
 0 & 0 & 0 & 0 & 0 & -1 \\
\end{array}
\right)\;,
\end{equation}
and by the lattice $\Lambda=\{e_1,\ldots,e_6\}$. 

On the other hand, the group $\langle F\rangle$ is generated by two 
roto--translations\footnote{Note that $\omega$ and $\vartheta\omega\vartheta$ 
belong to the same conjugacy class of $D_4$.} 
\begin{equation}
h_1=(\omega, 0) \quad\text{and}\quad h_2= g_1 g_2 g_1 = (\vartheta\omega\vartheta, \tfrac{1}{2}(e_1+e_3+e_5))
\end{equation}
and six translations $(\Id,e_i)$ for $i=2,4,5,6$, $(\Id,e_1+e_3)$ and 
$(\Id,-e_1+e_3)$, which define a (six--dimensional) sublattice 
$\Lambda_F\subset\Lambda$.

As a subgroup of $G$ the roto--translations $h_1$ and $h_2$ generate 
$G_F=\Z{2}\times\Z{2}$ and one can show that $D_4/\left(\Z{2}\times\Z{2}\right) = \Z{2}$, 
which is generated by $g_1$ using $g_1^2 = (\Id,e_1) \sim (\Id, 0)$ in the 
orbifolding group. Furthermore, one can take the quotient of the respective 
lattices and obtains $\Lambda/\Lambda_F = \Z{2}$, which is generated by 
$(\Id,e_3)$ using $(\Id,e_3) (\Id,e_3) = (\Id,2e_3) \sim (\Id,0)$ 
(or equivalently generated by $(\Id,e_1)$ using $(\Id,e_1) (\Id,e_1) 
= (\Id,2e_1) \sim (\Id,0)$). 

The full fundamental group $\pi_1 = S/\langle F\rangle$ of the orbifold 
$D_4$--1--5 is generated by $g_1$. Then, $g_1^2 = (\Id,e_1)$ 
(not identified with $(\Id,0)$ in $\pi_1$), 
$g_1^3=(\vartheta,\frac{3}{2} e_1+\frac{1}{4} e_5)$ and 
$g_1^4 = (\Id, 2e_1) \sim (\Id,0)$. Thus, we find $\pi_1 = \Z{4}$, see 
\Tabref{tab:FundamentalGroups}.

% ==============================================================================
\section{Examples}
\label{sec:examples}

In this section we discuss three examples of orbifolds with non--Abelian point 
group in detail. The first example in \Secref{sec:S3Example} considers $S_3$ 
\cite{Konopka:2012gy}, the easiest non--Abelian case, which unfortunately yields only
non--chiral spectra. Then, in \Secref{sec:T7Example} we discuss a $T_7$ orbifold 
which yields chirality. Furthermore, this model has the interesting property of 
having just one untwisted K\"ahler modulus and no untwisted complex structure 
modulus. Last, in \Secref{sec:Delta27Example} we describe a $\Delta(27)$ 
orbifold which possesses a non--trivial fundamental group and gives 
chirality.

\subsection{The heterotic $S_3$ orbifold}
\label{sec:S3Example}
The symmetric group $S_3$ is generated by two generators $\vartheta$ and $\omega$ 
of orders 2 and 3, i.e. $\vartheta^2=\omega^3=\Id$. They fulfill the 
relation $\vartheta\omega\vartheta=\omega^2$. $S_3$ has $3! = 6$ elements which split 
into three conjugacy classes as follows:
\begin{equation}
\label{eqn:S3ConjugacyClasses}
\begin{array}{lclcl}
  \left[\Id\right]      & = & \{\Id\}\;,                            &\qquad& v_{\left[\Id\right]}     =  \left(0,0,0\right)\;, \\
  \left[\omega\right]   & = & \{\omega, \omega^2\}\;,                     && v_{\left[\omega\right]}  =  \left(\frac{1}{3},-\frac{1}{3},0\right)\;, \\
  \left[\vartheta\right]   & = & \{\vartheta, \vartheta\omega, \vartheta\omega^2\}\,, && v_{\left[\vartheta\right]}  =  \left(\frac{1}{2},-\frac{1}{2},0\right)\;,
\end{array}
\end{equation}
where we listed for later use the corresponding twist vectors related to the corresponding \SU3--compatible
point--group generators given below in~\Eqref{eqn:S3asSU3} (obtained by choosing appropriate bases that diagonalize 
the respective rotation matrices).

From crystallography \cite{Plesken:1998,Fischer:2012qj}, we know that for this 
\QQ--class (i.e. point group $P=S_3$), there are six \ZZ--classes (i.e. 
inequivalent lattices) and in total eleven affine classes (i.e. for each lattice 
except for lattice \#6 there are two affine classes: first the trivial affine 
class without roto--translations and a second affine class where $g_\omega$ is a 
roto--translation), see \Tabref{tab:NonAbelianPointGroups}.

Let us discuss the first affine class, i.e. $S_3$--1--1. In this case the 
generators of the $S_3$ orbifolding group are 
$g_\vartheta = \left(\vartheta, 0\right)$ and $g_\omega = \left(\omega, 0\right)$, 
where
\begin{equation}
\vartheta_\basis{e} ~=~ 
\left(
\begin{array}{cccccc}
 1 & -1 & 0 & 0 & 0 & 0 \\
 0 & -1 & 0 & 0 & 0 & 0 \\
 0 & 0 & 0 & -1 & 0 & 0 \\
 0 & 0 & -1 & 0 & 0 & 0 \\
 0 & 0 & 0 & 0 & -1 & 0 \\
 0 & 0 & 0 & 0 & 0 & -1 \\
\end{array}
\right) \quad\text{and}\quad
\omega_\basis{e} ~=~ 
\left(
\begin{array}{cccccc}
 -1 & 1 & 0 & 0 & 0 & 0 \\
 -1 & 0 & 0 & 0 & 0 & 0 \\
 0 & 0 & 0 & 1 & 0 & 0 \\
 0 & 0 & -1 & -1 & 0 & 0 \\
 0 & 0 & 0 & 0 & 1 & 0 \\
 0 & 0 & 0 & 0 & 0 & 1 \\
\end{array}
\right)\;,
\end{equation}
given in the lattice basis as matrices from $\GL{6,\ZZ}$, for example, 
$\vartheta_\basis{e} e_1 = e_1$. One can go to the $\SO{6}$ form by a 
basis change $\vartheta = \bmat_\basis{e}\, \vartheta_\basis{e}\, \bmat_\basis{e}^{-1}$ 
and $\omega = \bmat_\basis{e}\, \omega_\basis{e}\, \bmat_\basis{e}^{-1}$, where 
the columns of $\bmat_{\basis{e}}$ are the basis vectors $e_i$, $i=1,\ldots,6$.
In the $\SU{3}$ basis these generators read 
(see Table C.2 of \cite{Fischer:2012qj})
\begin{equation}
\label{eqn:S3asSU3}
\vartheta^{(\rep{3})} ~=~ \left(
\begin{array}{ccc}
-1 & 0 & 0 \\
 0 & 0 & 1 \\
 0 & 1 & 0
\end{array}
\right) \quad\text{and}\quad
\omega^{(\rep{3})} ~=~ \left(
\begin{array}{ccc}
 1 & 0 & 0 \\
 0 & \mathrm{e}^{-2\pi\,\I\,\frac{1}{3}} & 0 \\
 0 & 0 & \mathrm{e}^{2\pi\,\I\,\frac{1}{3}}
\end{array}
\right)\;.
\end{equation}
These matrices generate a reducible three--dimensional representation 
$\rep{3}$ of $S_3$, which decomposes into irreducible representations as 
$\rep{3}=\rep{2}\oplus\rep{1'}$. Furthermore, there exists one 
additional irreducible representation of $S_3$: $\rep{1}$, the trivial 
singlet.

As discussed in \Secref{sec:hodgeNumbers}, the number of untwisted K\"ahler 
and complex structure moduli is determined by the tensor products of the 
three--dimensional representation of \Eqref{eqn:S3asSU3}, i.e.
\begin{subequations}
\begin{eqnarray}
\label{sec:S3tensor1}
\rep{3} \otimes \crep{3} & = & \left(\rep{2}\oplus\rep{1'}\right)\otimes\left(\crep{2}\oplus\crep{1'}\right) \; \rightarrow \rep{2}\oplus\rep{2}\oplus\rep{2}\oplus\rep{1'}\oplus\rep{1}\oplus\rep{1}\,,\\
\label{sec:S3tensor2}
\rep{3} \otimes \rep{3}  & = & \left(\rep{2}\oplus\rep{1'}\right)\otimes\left(\rep{2}\oplus\rep{1'}\right) \;   \rightarrow \rep{2}\oplus\rep{2}\oplus\rep{2}\oplus\rep{1'}\oplus\rep{1}\oplus\rep{1}\;.
\end{eqnarray}
\end{subequations}
Since \Eqref{sec:S3tensor1} contains two trivial singlets $\rep{1}$, there are two 
orbifold--invariant untwisted K\"ahler moduli from the states given in~\Eqref{eqn:UntwistedKahlerModulus}. 
Further, also \Eqref{sec:S3tensor2} contains two singlets $\rep{1}$ and 
hence there are also two orbifold--invariant untwisted complex structure moduli
from \Eqref{eqn:UntwistedComplexStructureModulus}. In total, we find
\begin{equation}
(h_\text{U}^{(1,1)}, h_\text{U}^{(2,1)}) = (2,2)\;.
\end{equation}

Next, we discuss the contributions from the two twisted sectors of the $S_3$ 
orbifold, specified by the inequivalent conjugacy classes given in~\Eqref{eqn:S3ConjugacyClasses}. 
The $[\omega]$ twisted sector has 
nine inequivalent constructing elements $g^{(i)} \in S$, $i=1,\ldots,9$ on the 
torus,
\begin{equation}
\begin{array}{lll}
g^{(1)} = \left(\omega, 0\right)       &,\;\; g^{(2)} = \left(\omega, e_4\right)         &,\;\; g^{(3)} = \left(\omega, 2e_4\right)\;,\\
g^{(4)} = \left(\omega, e_2\right)     &,\;\; g^{(5)} = \left(\omega, e_2+e_4\right)     &,\;\; g^{(6)} = \left(\omega, e_2+2e_4\right)\;, \\
g^{(7)} = \left(\omega, e_1+e_2\right) &,\;\; g^{(8)} = \left(\omega, e_1+e_2+e_4\right) &,\;\; g^{(9)} = \left(\omega, e_1+e_2+2e_4\right)\;. \\
\end{array}
\end{equation}
These constructing elements by themselves lead to a six--dimensional $\mathcal N=1$ supersymmetric theory, where the six dimensions
include the uncompactified 4D space along with the two--torus defined by the basis vectors $e_5$ and $e_6$.

Finally, the $[\vartheta]$ sector has four inequivalent constructing elements on the torus,
\begin{equation}
\left(\vartheta, n_5 e_5 + n_6 e_6\right) \quad\text{with}\quad n_5,n_6 = 0,1\;,
\end{equation}
which are also inequivalent on the orbifold.
As the $[\omega]$ sector, the $[\vartheta]$ twisted sector yields an $\mathcal N=1$ supersymmetric 
theory in the six dimensions composed of the uncompactified 4D space and the two--torus defined 
by the basis vectors $e_1$ and $e_4-e_3$.

The centralizer elements of the constructing elements of both twisted sectors do not further break supersymmetry 
in their respective six--dimensional $\mathcal N=1$ theories. Therefore, all $9+4$ fixed tori are endowed 
with both a $\rep{27}$-- and a $\crep{27}$--plet in four dimensions, 
contributing with as many twisted K\"ahler and complex--structure
moduli as the number of inequivalent constructing elements.

In summary, the Hodge numbers are $(h^{(1,1)}, h^{(2,1)}) = (15,15)$ arising from the various sectors as
\begin{equation}
(2, 2) U + (9, 9) T_{[\omega]} + (4, 4) T_{[\vartheta]}\;,
\end{equation}
confirming the results of \cite{Konopka:2012gy}. Unfortunately, in the standard 
heterotic CFT description the $S_3$ orbifold necessarily leads to a non--chiral 
spectrum in 4D, as we can see from the Hodge numbers $h^{(1,1)}=h^{(2,1)}$. 
Hence, the $S_3$ orbifold seems phenomenologically not promising. It might 
be possible to circumvent this by introducing magnetized tori 
\cite{Nibbelink:2012de}.

\subsection{The heterotic $T_7$ orbifold}
\label{sec:T7Example}
The Frobenius group $T_7$ is generated by two generators $\vartheta$ and $\omega$ 
of orders 3 and 7, i.e. $\vartheta^3=\omega^7=\Id$. They fulfill the 
relation $\omega\vartheta=\vartheta\omega^2$. $T_7$ has 21 elements, they split into 
five conjugacy classes, i.e.
\begin{equation}
\label{eqn:T7ConjugacyClasses}
\begin{array}{lcllcl}
  \left[\Id\right]      & = & \{\Id\}\;,                                                                                                  & v_{\left[\Id\right]}      & = & \left(0,0,0\right)\;, \\
  \left[\omega\right]   & = & \{\omega, \omega^2, \omega^4\}\;,                                                                           & v_{\left[\omega\right]}   & = & \left(\tfrac17,\tfrac27,-\tfrac37\right)\;, \\
  \left[\omega^3\right] & = & \{\omega^3, \omega^5, \omega^6\}\;,                                                                         & v_{\left[\omega^3\right]} & = & \left(-\tfrac17,-\tfrac27,\tfrac37\right)\;, \\
  \left[\vartheta\right]   & = & \{\vartheta, \vartheta\omega, \vartheta\omega^2, \vartheta\omega^3, \vartheta\omega^4, \vartheta\omega^5, \vartheta\omega^6\}\;, & v_{\left[\vartheta\right]}   & = & \left(\tfrac13,-\tfrac13,0\right)\;, \\
  \left[\vartheta^2\right] & = & \{\vartheta^2, \vartheta^2\omega, \vartheta^2 \omega^2, \vartheta^2 \omega^3, \vartheta^2 \omega^4, \vartheta^2 \omega^5, \vartheta^2 \omega^6\}\;, & v_{\left[\vartheta^2\right]} & = & \left(\tfrac13,-\tfrac13,0\right)\;. \\
\end{array}
\end{equation}
where, as in the $S_3$ example, we provide for later use the corresponding 
twist vectors associated with~\Eqref{eqn:T7asSU3}.

From crystallography \cite{Plesken:1998,Fischer:2012qj} we know that for this 
\QQ--class (i.e. point group $P=T_7$), there are three \ZZ--classes (i.e. 
inequivalent lattices) and in total three affine classes (i.e. for each lattice 
there is only the trivial affine class without roto--translations), see 
\Tabref{tab:NonAbelianPointGroups}.

Let us discuss the first \ZZ--class, i.e. $T_7$--1--1. In this case the 
generators of the $T_7$ orbifolding group are 
$g_\vartheta = \left(\vartheta, 0\right)$ and $g_\omega = \left(\omega, 0\right)$, 
where
\begin{equation}
\vartheta_\basis{e} ~=~ \left(\begin{array}{cccccc}
 0 &-1 & 0 & 0 &-1 & 0 \\
 0 &-1 & 0 & 1 &-1 & 0 \\
 0 & 1 & 0 &-1 & 1 &-1 \\
-1 & 0 & 0 & 0 & 0 &-1 \\
 0 & 0 &-1 & 0 & 1 & 0 \\
 1 & 0 & 0 & 0 & 1 & 0
\end{array}\right) \quad\text{and}\quad
\omega_\basis{e} ~=~ \left(\begin{array}{cccccc}
 0 &-1 & 0 & 0 & 0 &-1 \\
 1 &-1 & 0 & 0 & 0 &-1 \\
-1 & 1 & 0 & 0 &-1 & 1 \\
 1 & 0 & 1 & 0 &-1 & 0 \\
-1 & 1 & 0 & 0 & 0 & 0 \\
-1 & 1 & 0 &-1 & 0 & 0
\end{array}\right)\;,
\end{equation}
given in the lattice basis as matrices from $\GL{6,\ZZ}$, for example, 
$\vartheta_\basis{e} e_1 = -e_4+e_6$. One can go to the $\SO{6}$ form by a 
basis change $\vartheta = \bmat_\basis{e}\, \vartheta_\basis{e}\, \bmat_\basis{e}^{-1}$ 
and $\omega = \bmat_\basis{e}\, \omega_\basis{e}\, \bmat_\basis{e}^{-1}$, where 
the columns of $\bmat_{\basis{e}}$ are the basis vectors $e_i$, $i=1,\ldots,6$.
In the $\SU{3}$ basis these generators read 
(see Table C.2 of \cite{Fischer:2012qj})
\begin{equation}
\label{eqn:T7asSU3}
\vartheta^{(\rep{3})} ~=~ \left(
\begin{array}{ccc}
 0 & 1 & 0 \\
 0 & 0 & 1 \\
 1 & 0 & 0
\end{array}
\right) \quad\text{and}\quad
\omega^{(\rep{3})} ~=~ \left(
\begin{array}{ccc}
 \mathrm{e}^{2\pi\,\I\,\frac{4}{7}} & 0 & 0 \\
 0 & \mathrm{e}^{2\pi\,\I\,\frac{2}{7}} & 0 \\
 0 & 0 & \mathrm{e}^{2\pi\,\I\,\frac{1}{7}}
\end{array}
\right)\;.
\end{equation}
These matrices generate an irreducible three--dimensional representation 
$\rep{3}$ of $T_7$. Furthermore, there exist four additional 
irreducible representations of $T_7$: $\crep{3}$ is the complex conjugate of 
$\rep{3}$, $\rep{1'}$ and its complex conjugate $\crep{1'}$ are two 
non--trivial one--dimensional representations and, finally, $\rep{1}$ is the 
trivial singlet.

One can think of the $T_7$ orbifold as a standard $\Z{7}$ orbifold generated by 
$\omega$ with an additional, non--freely acting $\Z{3}$ generated by $\vartheta$ 
that permutes the three complex planes $(z_1,z_2,z_3)$ as $z_1 \mapsto z_3 
\mapsto z_2 \mapsto z_1$.

As discussed in \Secref{sec:hodgeNumbers}, the number of untwisted K\"ahler 
and complex structure moduli is determined by the tensor products of the 
three--dimensional representation of \Eqref{eqn:T7asSU3}, i.e.
\begin{subequations}
\begin{eqnarray}
\label{sec:T7tensor1}
\rep{3} \otimes \crep{3} & \rightarrow & \rep{3}\oplus\crep{3}\oplus\rep{1}\oplus\rep{1'}\oplus\crep{1'}\,,\\
\label{sec:T7tensor2}
\rep{3} \otimes \rep{3}  & \rightarrow & \left(\rep{3}\oplus\crep{3}\right)_s \oplus \crep{3}_a\;,
\end{eqnarray}
\end{subequations}
where $s$ and $a$ denotes the symmetric and anti--symmetric part, respectively. 
As \Eqref{sec:T7tensor1} contains one singlet $\rep{1}$, there is one 
orbifold--invariant untwisted K\"ahler modulus from \Eqref{eqn:UntwistedKahlerModulus}. 
Furthermore, \Eqref{sec:T7tensor2} does not contain the singlet $\rep{1}$ and 
hence there is no orbifold--invariant untwisted complex structure modulus 
from \Eqref{eqn:UntwistedComplexStructureModulus}. In summary, we find
\begin{equation}
(h_\text{U}^{(1,1)}, h_\text{U}^{(2,1)}) = (1,0)\;.
\end{equation}

Next, we study the contributions from the four twisted sectors of the $T_7$ 
orbifold arising from its conjugacy classes (see \Eqref{eqn:T7ConjugacyClasses}). 
The $[\omega]$ twisted sector has seven inequivalent constructing elements 
$g^{(i)} \in S$, $i=1,\ldots,7$, on the torus:
\begin{equation}
\begin{array}{ll}
g^{(1)} = \left(\omega, 0\right)                     &,\;\; g^{(2)} = \left(\omega, e_1+e_2\right)        \;, \\
g^{(3)} = \left(\omega, e_1+e_2+e_6\right)           &,\;\; g^{(4)} = \left(\omega, e_1+e_3+e_5+e_6\right)\;, \\
g^{(5)} = \left(\omega, 2e_1+2e_2+e_6\right)         &,\;\; g^{(6)} = \left(\omega, 2e_1+e_2+e_3+e_5+e_6\right)\;, \\
g^{(7)} = \left(\omega, 2e_1+e_2+e_3+e_5+2e_6\right) &. 
\end{array}
\end{equation}
They are also inequivalent on the orbifold. The corresponding fixed points are given in the $e_\alpha$ basis by
$f^{(i)}= \frac{1}{7} \hat{f}^{(i)}_\alpha e_\alpha$, $i=1,\ldots,7$, with
\begin{equation}
\begin{array}{lll}
\hat{f}^{(1)} = \left(0, 0, 0, 0, 0, 0\right) &,\;\; \hat{f}^{(2)} = \left(2, 4, 1, 1, 2, 1\right) &,\;\; \hat{f}^{(3)} = \left(1, 2, 4, 4, 1, 4\right)\;, \\
\hat{f}^{(4)} = \left(4, 1, 2, 2, 4, 2\right) &,\;\; \hat{f}^{(5)} = \left(3, 6, 5, 5, 3, 5\right) &,\;\; \hat{f}^{(6)} = \left(6, 5, 3, 3, 6, 3\right)\;, \\
\hat{f}^{(7)} = \left(5, 3, 6, 6, 5, 6\right)\;. & &
\end{array}
\end{equation}
As these are fixed points (and not tori) and the centralizers of $g^i$ are 
trivial, the $[\omega]$ twisted sector combines with the inverse twisted sector 
$[\omega^6] = [\omega^3]$ and gives seven left--chiral $\rep{27}$--plets plus 
their CPT conjugate partners. Hence, this sector contributes with $(7,0)$ to 
the Hodge numbers. 

The $[\vartheta]$ twisted sector has one inequivalent constructing element with 
associated fixed torus,
\begin{equation}
\left(\vartheta, 0\right) \quad\text{with}\quad f = \left(f_1, f_2, 0, -f_1+f_2, -f_1-f_2, -f_2\right)\;,
\end{equation}
where the torus is parametrized by $f_1, f_2 \in \RR$. As the centralizer of 
this sector is trivial, the $[\vartheta]$ twisted sector feels the full 
$\mathcal{N}=1$ in six dimensions and hence contributes $(1,1)$ to the Hodge 
numbers.

Finally, the $[\vartheta^2]$ is very similar to the $[\vartheta]$ twisted sector. It 
has one inequivalent constructing element with associated fixed torus,
\begin{equation}
\left(\vartheta^2, 0\right) \quad\text{with}\quad f = \left(f_1, f_2, 0, -f_1+f_2, -f_1-f_2, -f_2\right)\;,
\end{equation}
where the torus is parametrized by $f_1, f_2 \in \RR$. Again, as the centralizer 
is trivial, it gives rise to one twisted K\"ahler and one twisted complex 
structure modulus and therefore contributes $(1,1)$ to the Hodge numbers.

In summary, the Hodge numbers are $(h^{(1,1)}, h^{(2,1)}) = (10,2)$, distributed 
in the various sectors according to
\begin{equation}
(1, 0) U + (7, 0) T_{[\omega]} + (1, 1) T_{[\vartheta]} + (1, 1) T_{[\vartheta^{2}]}\;.
\end{equation}

\subsection{The heterotic $\Delta(27)$ orbifold}
\label{sec:Delta27Example}
The group $\Delta(27)$ is generated by two generators $\vartheta$ and $\omega$ 
both of order 3, i.e. such that $\vartheta^3=\omega^3=\Id$. $\Delta(27)$ has 27 
elements, they split into the following eleven conjugacy classes
\begin{equation}
\label{eqn:Delta27ConjugacyClasses}
\begin{array}{lclcll}
  \left[\Id\right]                          & = & \{\Id\}\;,                                                    &\qquad& v_{\left[\Id\right]}                          &\!\!\!\! = \left(0,0,0\right)\;, \\
  \left[\omega\right]                       & = & \{\omega, \vartheta\omega\vartheta^2, \vartheta\omega^2\vartheta^2\omega^2\}\;, && v_{\left[\omega\right]}                       &\!\!\!\! =  \left(\frac{1}{3},-\frac{1}{3},0\right)\;, \\
  \left[\omega^2\right]                     & = & \{\omega^2, \vartheta\omega^2\vartheta^2, \vartheta\omega\vartheta^2\omega\}\;, && v_{\left[\omega^2\right]}                     &\!\!\!\! =  \left(\frac{1}{3},-\frac{1}{3},0\right)\;, \\
  \left[\vartheta\right]                       & = & \{\vartheta, \omega\vartheta\omega^2, \omega^2\vartheta\omega\}\;,           && v_{\left[\vartheta\right]}                       &\!\!\!\! =  \left(\frac{1}{3},-\frac{1}{3},0\right)\;, \\
  \left[\vartheta^2\right]                     & = & \{\vartheta^2, \omega^2\vartheta^2\omega, \omega\vartheta^2\omega^2\}\;,     && v_{\left[\vartheta^2\right]}                     &\!\!\!\! =  \left(\frac{1}{3},-\frac{1}{3},0\right)\;, \\
  \left[\omega\vartheta\right]                 & = & \{\omega\vartheta, \omega^2\vartheta\omega^2, \vartheta\omega\}\;,           && v_{\left[\omega\vartheta\right]}                 &\!\!\!\! =  \left(\frac{1}{3},-\frac{1}{3},0\right)\;, \\
  \left[\omega\vartheta^2\right]               & = & \{\omega\vartheta^2, \vartheta^2\omega, \omega^2\vartheta^2\omega^2\}\;,     && v_{\left[\omega\vartheta^2\right]}               &\!\!\!\! =  \left(\frac{1}{3},-\frac{1}{3},0\right)\;, \\
  \left[\omega^2\vartheta\right]               & = & \{\omega^2\vartheta, \vartheta\omega^2, \omega\vartheta\omega\}\;,           && v_{\left[\omega^2\vartheta\right]}               &\!\!\!\! =  \left(\frac{1}{3},-\frac{1}{3},0\right)\;, \\
  \left[\omega^2\vartheta^2\right]             & = & \{\omega^2\vartheta^2, \omega\vartheta^2\omega, \vartheta^2\omega^2\}\;,     && v_{\left[\omega^2\vartheta^2\right]}             &\!\!\!\! =  \left(\frac{1}{3},-\frac{1}{3},0\right)\;, \\
  \left[\vartheta\omega\vartheta^2\omega^2\right] & = & \{\vartheta\omega\vartheta^2\omega^2\}\;,                                 && v_{\left[\vartheta\omega\vartheta^2\omega^2\right]} &\!\!\!\! =  \left(\frac{1}{3},\frac{1}{3},-\frac{2}{3}\right)\;, \\
  \left[\vartheta\omega^2\vartheta^2\omega\right] & = & \{\vartheta\omega^2\vartheta^2\omega\}\;,                                 && v_{\left[\vartheta\omega^2\vartheta^2\omega\right]} &\!\!\!\! =  \left(\frac{1}{3},\frac{1}{3},-\frac{2}{3}\right)\;, \\
\end{array}
\end{equation}
where we also give the corresponding twist vectors obtained, as before, by 
choosing bases in which the rotation matrices are diagonal, as 
in~\Eqref{eqn:Delta27asSU3}.

Once again, it is known that there are three lattices and a total of ten affine 
classes (three orbifolding groups without roto--translations and seven ones 
which include them) for the point group $P=\Delta(27)$.

Let us discuss the fourth affine class of the first \ZZ--class, i.e. 
$\Delta(27)$--1--4, see \Tabref{tab:NonAbelianPointGroups}. In this case the 
generators of the $\Delta(27)$ orbifolding group are 
$g_\vartheta = (\vartheta,\frac{1}{3}(2e_2+e_3+2e_5))$ and 
$g_\omega = (\omega,\frac{1}{3} e_1)$, where
\begin{equation}
\vartheta_\basis{e} ~=~ \left(\begin{array}{cccccc}
 0 & 1 & 0 & 0 & 1 & 0 \\
 0 & 0 & 0 & 0 & 0 & 1 \\
 0 & 1 & 0 &-1 & 2 & 1 \\ 
 0 & 0 & 1 &-1 & 1 & 1 \\ 
 0 & 0 & 0 & 0 & 1 & 0 \\ 
 1 & 0 & 0 & 0 &-1 & 0
\end{array}\right) \quad\text{and}\quad
\omega_\basis{e} ~=~ \left(\begin{array}{cccccc}
 1 & 1 &-1 & 0 & 0 &-1 \\
 0 &-1 & 1 &-1 & 0 & 1 \\
 0 & 0 & 0 &-1 & 0 &-1 \\
 0 & 0 & 0 & 0 &-1 &-1 \\
 0 & 1 & 0 & 0 & 0 &-1 \\
 0 &-1 & 0 & 1 &-1 & 0
\end{array}\right)\;,
\end{equation}
given in the lattice basis as matrices from $\GL{6,\ZZ}$, for example, 
$\vartheta_\basis{e} e_1 = e_6$. One can go to the $\SO{6}$ form by a 
basis change $\vartheta = \bmat_\basis{e}\, \vartheta_\basis{e}\, \bmat_\basis{e}^{-1}$ 
and $\omega = \bmat_\basis{e}\, \omega_\basis{e}\, \bmat_\basis{e}^{-1}$, where 
the columns of $\bmat_{\basis{e}}$ are the basis vectors $e_i$, $i=1,\ldots,6$.
In the $\SU{3}$ basis these generators read 
(see Table C.2 of \cite{Fischer:2012qj})
\begin{equation}
\label{eqn:Delta27asSU3}
\vartheta^{(\rep{3})} ~=~ \left(
\begin{array}{ccc}
 0 & 1 & 0 \\
 0 & 0 & 1 \\
 1 & 0 & 0
\end{array}
\right) \quad\text{and}\quad
\omega^{(\rep{3})} ~=~ \left(
\begin{array}{ccc}
  1 &                                  0 &                                 0 \\
  0 & \mathrm{e}^{2\pi\,\I\,\frac{1}{3}} &                                 0 \\
  0 &                                  0 & \mathrm{e}^{2\pi\,\I\,\frac{2}{3}}
\end{array}
\right)\;,
\end{equation}
which generate an irreducible three--dimensional representation $\rep3$ of $\Delta(27)$.

The number of untwisted moduli corresponds to the number of invariant singlets
within the tensor products of the three--dimensional representation and its conjugate:
\begin{subequations}
\begin{eqnarray}
\label{sec:Delta27tensor1}
\rep{3} \otimes \crep{3} & \rightarrow & \rep{1_0}\oplus\rep{1_1}\oplus\rep{1_2}\oplus\rep{1_3}\oplus\rep{1_4}\oplus\rep{1_5}\oplus\rep{1_6}\oplus\rep{1_7}\oplus\rep{1_8}\,,\\
\label{sec:Delta27tensor2}
\rep{3} \otimes \rep{3}  & \rightarrow & \crep{3}\oplus\crep{3}\oplus\crep{3}\;,
\end{eqnarray}
\end{subequations}
where only $\rep{1_0}$ in~\Eqref{sec:Delta27tensor1} denotes a 
$\Delta(27)$--invariant singlet. Therefore, by using~\Eqref{eq:TensorProduct}, 
we conclude that there is only one orbifold--invariant untwisted K\"ahler 
modulus and no orbifold--invariant untwisted complex--structure modulus, i.e.
\begin{equation}
(h_\text{U}^{(1,1)}, h_\text{U}^{(2,1)}) = (1,0)\;.
\end{equation}

The only nonvanishing contributions to the Hodge numbers from the twisted 
sectors arise from the 27 fixed points of the 
$T_{[\vartheta\omega\vartheta^{2}\omega^{2}]}$ sector, which are inequivalent 
on the torus. The constructing elements associated to these fixed points are 
$g^{(i)} = \left(\vartheta\omega\vartheta^{2}\omega^{2},\lambda^{(i)}\right)$ 
with:
\begin{equation}
\label{eqn:Delta27ConstrcutingElements}
\mbox{\footnotesize$
\begin{array}{ll}
\lambda^{(1)}  = \tfrac13(-2e_1+5e_2+e_3-e_4-e_5+5e_6)\,,   & \lambda^{(2)}  = \tfrac13(-2e_1+5e_2+e_3+2e_4-e_5+5e_6) \,,\\
\lambda^{(3)}  = \tfrac13(e_1+2e_2+e_3-e_4-e_5+2e_6)\,,     & \lambda^{(4)}  = \tfrac13(e_1+2e_2+e_3+2e_4-e_5+2e_6)\,,\\
\lambda^{(5)}  = \tfrac13(e_1+2e_2+e_3+5e_4-e_5+2e_6)\,,    & \lambda^{(6)}  = \tfrac13(e_1+5e_2+e_3-e_4-e_5+2e_6)\,, \\
\lambda^{(7)}  = \tfrac13(e_1+5e_2+e_3+2e_4-e_5+2e_6)\,,    & \lambda^{(8)}  = \tfrac13(e_1+5e_2+e_3+5e_4-e_5+2e_6)\,,\\
\lambda^{(9)}  = \tfrac13(-2e_1+5e_2-2e_3-e_4-4e_5+5e_6)\,, & \lambda^{(10)} = \tfrac13(-2e_1+5e_2-2e_3-e_4-e_5+5e_6)\,,\\
\lambda^{(11)} = \tfrac13(-2e_1+5e_2+e_3-e_4-4e_5+5e_6)\,,  & \lambda^{(12)} = \tfrac13(-2e_1+5e_2+e_3+2e_4-4e_5+5e_6)\,,\\
\lambda^{(13)} = \tfrac13(-2e_1+8e_2-2e_3-e_4-4e_5+5e_6)\,, & \lambda^{(14)} = \tfrac13(-2e_1+8e_2+e_3-e_4-4e_5+5e_6)\,,\\
\lambda^{(15)} = \tfrac13(-2e_1+8e_2+e_3+2e_4-4e_5+5e_6)\,, & \lambda^{(16)} = \tfrac13(e_1+2e_2-2e_3-e_4-4e_5+5e_6)\,,\\
\lambda^{(17)} = \tfrac13(e_1+2e_2-2e_3+2e_4-4e_5+5e_6)\,,  & \lambda^{(18)} = \tfrac13(e_1+2e_2+e_3+2e_4-4e_5+5e_6)\,,\\
\lambda^{(19)} = \tfrac13(e_1+5e_2-2e_3-e_4-4e_5+5e_6)\,,   & \lambda^{(20)} = \tfrac13(e_1+5e_2-2e_3+2e_4-4e_5+5e_6)\,,\\
\lambda^{(21)} = \tfrac13(e_1+5e_2+e_3-e_4-4e_5+2e_6)\,,    & \lambda^{(22)} = \tfrac13(e_1+5e_2+e_3+2e_4-4e_5+2e_6)\,,\\
\lambda^{(23)} = \tfrac13(e_1+5e_2+e_3+2e_4-4e_5+5e_6)\,,   & \lambda^{(24)} = \tfrac13(e_1+5e_2+e_3+5e_4-4e_5+2e_6)\,,\\
\lambda^{(25)} = \tfrac13(e_1+8e_2-2e_3-e_4-4e_5+5e_6)\,,   & \lambda^{(26)} = \tfrac13(e_1+8e_2-2e_3+2e_4-4e_5+5e_6)\,,\\
\lambda^{(27)} = \tfrac13(e_1+8e_2+e_3+2e_4-4e_5+5e_6)\,.   & \\
\end{array}
$}
\end{equation}
Out of these 27 constructing elements, only three are inequivalent on the orbifold. We choose $g^{(1)}, g^{(2)}$ and $g^{(3)}$.
The corresponding fixed points are localized at
$f^{(i)}= \frac{1}{9} \hat{f}^{(i)}_\alpha e_\alpha$, $i=1,2,3$, with 
\begin{equation}
\begin{array}{lll}
\hat{f}^{(1)} = \left(1, 1, 6, 5, 6, 7\right)\,, &\hat{f}^{(2)} = \left(1, 1, 3, 8, 6, 7\right)\,, &\hat{f}^{(3)} = \left(4, 1, 6, 2, 3, 1\right)\;. 
\end{array}
\end{equation}

Since these are fixed points (and not tori) and the centralizers of $g^{(i)}$ are trivial, 
the $T_{[\vartheta\omega\vartheta^{2}\omega^{2}]}$ twisted sector combines with the inverse twisted sector 
$T_{[\vartheta\omega^2\vartheta^{2}\omega]}$ yielding three left--chiral $\rep{27}$--plets plus their 
CPT conjugate partners. Hence, the only twisted contribution to the Hodge numbers is $(3,0)$. 

In summary, the Hodge numbers are $(h^{(1,1)}, h^{(2,1)}) = (4,0)$ originating from the various sectors as
\begin{equation}
(1, 0) U+(3, 0) T_{[\vartheta\omega\vartheta^{2}\omega^{2}]}\;.
\end{equation}

The main feature that distinguishes this case from the previous examples
is the existence of a non--trivial fundamental group $\pi_1=S/\langle F\rangle$.
The group $\langle F\rangle$ generated by the set $F$ of the constructing elements
listed in \Eqref{eqn:Delta27ConstrcutingElements} contains the full lattice 
$\Lambda$ of the space group $S$ and a (normal subgroup) 
$\Z{3} \subset \Delta(27)$ generated by 
$\vartheta\omega\vartheta^2\omega^2$. Thus, we identify the fundamental group of the 
$\Delta(27)$--1--4 orbifold as
\begin{equation}
\pi_1 ~=~ S/\langle F\rangle ~=~ \Delta(27)/\Z{3} ~=~ \Z{3}\times\Z{3}\;.
\end{equation}

\vspace{-8mm}
% ==============================================================================
\section{Summary and Discussion}
\label{sec:Discussion}

We have computed systematically the number of (untwisted and twisted) moduli 
and fundamental groups of all 331 recently classified~\cite{Fischer:2012qj} 
$\mathcal N=1$ non--Abelian (symmetric) orbifold compactifications of the 
$\E{8}\times\E{8}'$ heterotic string with standard gauge embedding. We have 
developed the tools that allow us to determine the number of K\"ahler and 
complex--structure moduli by using group--theoretical and geometrical 
properties of the orbifolds rather than by direct computation. Our results are 
presented in~\Tabref{tab:NonAbelianPointGroups}, where the Hodge numbers, 
classified by sector, are displayed. Furthermore, we list all 38 non--trivial 
fundamental groups in~\Tabref{tab:FundamentalGroups}. Further details (such as 
orbifold generators, constructing elements, non--trivial centralizer elements, 
compactification lattices, etc.) are made available at\\
\phantom.\qquad\url{http://einrichtungen.physik.tu-muenchen.de/T30e/codes/NonAbelianOrbifolds/}\\
in a \texttt{Mathematica}--compatible format.

Most of the fundamental groups (35 out of 38) are Abelian 
(see~\Tabref{tab:FundamentalGroups}), such as $\Z{2}$, $\Z{3}$, $\Z{4}$ and 
$\Z{2}\times\Z{2}$\footnote{This is in contrast to smooth Calabi--Yau spaces, 
which have a much wider variety of fundamental groups, see e.g. 
\cite{Anderson:2009mh,Braun:2010vc} for fundamental groups of complete 
intersection Calabi--Yau threefolds.}. In 14 cases the fundamental group is 
generated by translations, in 16 cases all generators are rotations and in the 
remaining 8 cases the fundamental group is generated by translations and 
rotations. From a phenomenological point of view, orbifolds with non--trivial 
fundamental groups are very interesting as they may allow for non--local GUT 
breaking, which can improve gauge coupling unification. Furthermore, it would 
be interesting to study the connection of these orbifolds to smooth Calabi--Yau 
spaces \cite{Blaszczyk:2009in,Blaszczyk:2010db}, since the standard model gauge 
group (especially the hypercharge) can survive a full blow--up of the orbifold 
to a smooth Calabi--Yau when the fundamental group of the orbifold is 
non--trivial and the gauge group is broken non--locally.

Besides the fact that, like almost all Abelian cases, most non--Abelian orbifold 
geometries satisfy the relation $\chi=0\mod12$, for which we have no explanation, we observe that, 
in contrast to Abelian orbifolds, there is a large number of geometries (and a greater
number of models) with the overall volume modulus as the only untwisted modulus
available. These models should be further analyzed in the context of 
no--scale supergravity. Note also that this might be a positive feature for moduli stabilization, although
unfortunately it prevents anisotropic compactifications, which are desirable to
solve the tension between the string scale and the GUT scale~\cite{Witten:1996mz,Hebecker:2004ce}.

An interesting observation is that 42 out of the 331 orbifold geometries 
have vanishing Euler numbers $\chi$ (i.e. $h^{(1,1)} = h^{(2,1)}$). In these cases we note 
that we have $h_\text{U}^{(1,1)} = h_\text{U}^{(2,1)}$ and $h_\text{T}^{(1,1)} = 
h_\text{T}^{(2,1)}$, independently. The latter, $h_\text{T}^{(1,1)} = 
h_\text{T}^{(2,1)}$, is related to higher--dimensional supersymmetry. Hence, 
4D chiral spectra can never be obtained in these cases using standard heterotic 
orbifold CFT techniques alone. The inclusion of magnetized tori~\cite{Abe:2009uz,Nibbelink:2012de} 
may offer a plausible way to circumvent this hurdle.
However, their description is only known in blow--up, but not on the singular orbifold.

Furthermore, it would be interesting to analyze the cases of vanishing Euler 
numbers in the context of~\cite{KashaniPoor:2013en}, which states that type II 
string theory compactified on Calabi--Yau threefolds with vanishing Euler numbers leads to 
$\mathcal{N}=4$ enhanced supersymmetry (spontaneously broken to $\mathcal{N}=2$). 
Translated to the case of heterotic orbifolds with standard embedding and vanishing Euler numbers, one 
might expect $\mathcal{N}=2$ enhanced supersymmetry (spontaneously broken to 
$\mathcal{N}=1$). In addition, we find cases where $h_\text{T}^{(1,1)} = 
h_\text{T}^{(2,1)} = 0$, for example, $D_4-1-6$ has only untwisted Hodge 
numbers $(2,2)$ (see \Tabref{tab:NonAbelianPointGroups}) and there are few other similar 
cases in Abelian orbifolds~\cite{Fischer:2012qj}. One might conjecture that these cases give even 
higher enhanced supersymmetry, i.e. (spontaneously broken) $\mathcal{N}=4$. On 
the other hand, there are cases of orbifolds with vanishing Euler numbers where 
the orbifold intuition naively contradicts the general results of 
\cite{KashaniPoor:2013en}: e.g. consider the first case of 
\Tabref{tab:NonAbelianPointGroups}, $S_3$--1--1, with Hodge numbers $(15,15)$ 
decomposed as $(2, 2) U+(4, 4) T_{[\vartheta]}+(9, 9) T_{[\omega]}$. In this case, 
the two twisted sectors $T_{[\vartheta]}$ and $T_{[\omega]}$ both feel 
$\mathcal{N}=2$ supersymmetry, while the untwisted sector $U$ is $\mathcal{N}=1$
in four dimensions.
However, $T_{[\vartheta]}$ has different $\mathcal{N}=2$ than $T_{[\omega]}$,
as one can easily verify by noticing that the generators $\vartheta$ and $\omega$ 
leave untouched different two--tori. This implies that the full action of the orbifold
breaks explicitly (not spontaneously) $\mathcal{N}=2$ to $\mathcal{N}=1$ supersymmetry,
even though $\chi=0$.

In addition, we have presented the details of three sample models with the point 
groups $S_3$ (confirming the results of Ref.~\cite{Konopka:2012gy}), $T_7$
and $\Delta(27)$. We have chosen these point groups because they illustrate
the main properties of non--Abelian orbifold compactifications and because 
of their relevance in particle physics, for example in the context of neutrino mixing and 
family symmetries (see e.g.~\cite{Ishimori:2012zz,Luhn:2007sy,Luhn:2012bc}). 
As in the Abelian case, we expect the (non--Abelian) point group of the 
orbifold to be in close connection with the family symmetry of the 4D effective theory 
via string--selection rules~\cite{Kobayashi:2006wq}. If this was the case, our
examples would be of phenomenological interest. Yet the specifics of the string 
selection rules for non--Abelian orbifolds should still be worked out.

The results of this work lay the foundation stone of future phenomenological
studies based on non--Abelian orbifolds and can be extended in various ways.
Particularly, it would be interesting to extend this study to type IIA 
strings on orientifolds~\cite{Blumenhagen:1999ev,Forste:2000hx,Blumenhagen:2006ab},
where appealing phenomenology can also emerge.
Likewise, it might be desirable to apply our techniques to compactifications of
the heterotic strings on four--dimensional orbifolds, in order to reveal 
further connections to K3 manifolds~\cite{Erler:1993zy,Honecker:2006qz}.
Finally, one is now in position to tackle the technical details of the gauge embedding
in order to possibly arrive at promising constructions. In this respect, it is 
phenomenologically relevant to emphasize that in general the rank of the gauge 
group shall be reduced for non--Abelian orbifolds, which is in contrast to the 
situation in Abelian orbifolds, where the rank is always 16 after 
compactification. This can help avoiding multiple Higgs mechanisms 
to arrive at phenomenologically viable constructions from string theory. 

%\newpage
\subsection*{Acknowledgments}
We would like to thank Sebastian Konopka, Jan Louis and Michael Ratz for useful discussions.
The authors thank the Bethe Center for Theoretical Physics in Bonn, where part of this project was done, for hospitality and support.
P.V.\ is supported by SFB grant 676 and is grateful to the Institute of Physics, UNAM, for hospitality and support. 
S.R-S. is partially supported by CONACyT grant 151234 and DGAPA-PAPIIT grant IB101012. M.F.\ is supported by the Deutsche 
Forschungsgemeinschaft (DFG) through the cluster of excellence ``Origin and Structure of the Universe''.

\newpage
\appendix

% ==============================================================================
\begin{landscape}
\section{Results}
\label{sec:NonAbelianResults}

In this appendix we list the generators of the orbifolding group, the total 
Hodge numbers, their contributions from the various twisted and untwisted 
sectors and the mechanism of higher--dimensional gauge group breaking (local or 
non--local, see \Tabref{tab:FundamentalGroups}) for all 331 orbifolds 
with non--Abelian point group. For example, 
consider the $S_3$ point group with \ZZ--class \# 6 and affine class \# 1 
(i.e. no roto--translations and the orbifolding group is generated by 
$(\vartheta,0)$ and $(\omega,0)$). The higher--dimensional gauge group is broken 
locally in higher dimensions, which corresponds to a trivial fundamental group. 
The Hodge numbers are $(7,7)$, where $(2,2)$ originate from the untwisted sector $U$, 
$(4, 4)$ from the twisted sector $T_{[\vartheta]}$ and, finally, $(1, 1)$ from 
$T_{[\omega]}$.

{
\footnotesize

\begin{center}
\begin{longtable}{|c|c|c|l|c|}
\hline
\QQ--class ($P$), & \ZZ--         & affine           & generators of $G$ & \\
GAPID,            & \!\!class\!\! & class,           &                   & \\
\cline{4-4}
\Carat\ index     & ($\Lambda$)   & \!\!breaking\!\! & contributions to $(h^{(1,1)},h^{(2,1)})$ from $U$ and $T$ sectors & \!\!\!$(h^{(1,1)},h^{(2,1)})$\!\!\! \\
\hline
\hline
\endfirsthead
\hline
\QQ--class ($P$), & \ZZ--         & affine           & generators of $G$ & \\
GAPID,            & \!\!class\!\! & class,           &                   & \\
\cline{4-4}
\Carat\ index     & ($\Lambda$)   & \!\!breaking\!\! & contributions to $(h^{(1,1)},h^{(2,1)})$ from $U$ and $T$ sectors & \!\!\!$(h^{(1,1)},h^{(2,1)})$\!\!\! \\
\hline
\hline
\endhead
\hline
\multicolumn{5}{|r|}{continued \ldots}\\
\hline
\endfoot
\endlastfoot
\hline
$S_3$             & 1 &           1 & $(\vartheta, 0), (\omega, 0)$ & \\
\cline{4-4}
\GAP{6,1}         &   &       local & $(2, 2) U+(4, 4) T_{[\vartheta]}+(9, 9) T_{[\omega]}$ & $(15, 15)$ \\
\cline{3-5}
2262              &   &           2 & $(\vartheta, 0), (\omega,\frac{1}{3} e_5)$ & \\
\cline{4-4}
                  &   &  non--local & $(2, 2) U+(4, 4) T_{[\vartheta]}$ & $(6, 6)$ \\
\cline{2-5}
                  & 2 &           1 & $(\vartheta, 0), (\omega, 0)$ & \\
\cline{4-4}
                  &   &       local & $(2, 2) U+(4, 4) T_{[\vartheta]}+(9, 9) T_{[\omega]}$ & $(15, 15)$ \\
\cline{3-5}
                  &   &           2 & $(\vartheta, 0), (\omega,\frac{1}{3} e_5)$ & \\
\cline{4-4}
                  &   &  non--local & $(2, 2) U+(4, 4) T_{[\vartheta]}$ & $(6, 6)$ \\
\end{longtable}
\end{center}

}

\end{landscape}

\begin{table}[!hb]
\centering
{\small
\begin{tabular}{|c|c||c||cc||c|}
\hline
point group $P$ & \ZZ-- and    & Hodge numbers           & \multicolumn{2}{|c||}{origin of $\pi_1$ generators:} & fundamental group            \\
(\QQ--class)    & affine class & $(h^{(1,1)},h^{(2,1)})$ & $G/\langle G_F\rangle$ & $\Lambda/\Lambda_F$         & $\pi_1 = S/\langle F\rangle$ \\
\hline
\hline
$S_3$        &  1--2  &   $(6,6)$ & \Id                & $\Z{3}\times\Z{3}$ & $\Z{3}\times\Z{3}$ \\
             &  2--2  &   $(6,6)$ & \Id                & $\Z{3}$            & $\Z{3}$            \\
             &  3--2  &   $(6,6)$ & \Id                & $\Z{3}$            & $\Z{3}$            \\
\hline
$D_4$        &  1--3  & $(11,11)$ & $\Z{2}$            & $\Z{}^2$           & $\Z{2}\ltimes\Z{}^2$ \\
             &  1--5  &   $(6,6)$ & $\Z{2}$            & $\Z{2}$            & $\Z{4}$            \\
             &  1--6  &   $(2,2)$ & $D_4$              & $\Z{}^6$           & $S$                \\
\hline
\end{tabular}
}
\vspace{-0.1cm}
\caption{List of all non--trivial fundamental groups for orbifolds with 
non--Abelian $P$. The first column specifies $P$ and the second column 
enumerates the respective \ZZ-- and affine classes. In the third column we list 
the Hodge numbers in order to identify those cases which allow for chiral 
spectra, c.f. \cite{Nibbelink:2012de}. The forth and fifth column help to 
identify the origin of the generators of $\pi_1$ from the orbifolding group $G$ 
and from the lattice $\Lambda$, respectively. Finally, the last column lists $\pi_1$.}
\label{tab:FundamentalGroups}
\end{table}

\bibliography{Orbifold}

\providecommand{\bysame}{\leavevmode\hbox to3em{\hrulefill}\thinspace}
\frenchspacing
\newcommand{\origttfamily}{}
\let\origttfamily=\ttfamily
\renewcommand{\ttfamily}{\origttfamily \hyphenchar\font=`\-}

\begin{thebibliography}{10}

\bibitem{Dixon:1985jw}
L.~J. Dixon, J.~A. Harvey, C.~Vafa, and E.~Witten, Nucl.Phys. \textbf{B261}
  (1985), 678.
%%CITATION = NUPHA,B261,678;%%

\bibitem{Dixon:1986jc}
L.~J. Dixon, J.~A. Harvey, C.~Vafa, and E.~Witten, Nucl.Phys. \textbf{B274}
  (1986), 285.
%%CITATION = NUPHA,B274,285;%%

\bibitem{Faraggi:1991jr}
A.~E. Faraggi, Phys.Lett. \textbf{B278} (1992), 131.
%%CITATION = PHLTA,B278,131;%%

\bibitem{Dijkstra:2004cc}
T.~Dijkstra, L.~Huiszoon, and A.~Schellekens, Nucl.Phys. \textbf{B710} (2005),
  3, \texttt{arXiv:hep-th/0411129} [hep-th].
%%CITATION = HEP-TH/0411129;%%

\bibitem{Gmeiner:2005vz}
F.~Gmeiner, R.~Blumenhagen, G.~Honecker, D.~L{\"u}st, and T.~Weigand, JHEP
  \textbf{0601} (2006), 004, \texttt{arXiv:hep-th/0510170} [hep-th].
%%CITATION = HEP-TH/0510170;%%

\bibitem{Blumenhagen:2006ci}
R.~Blumenhagen, B.~K{\"o}rs, D.~L{\"u}st, and S.~Stieberger, Phys.Rept.
  \textbf{445} (2007), 1, \texttt{arXiv:hep-th/0610327} [hep-th].
%%CITATION = HEP-TH/0610327;%%

\bibitem{Gmeiner:2008xq}
F.~Gmeiner and G.~Honecker, JHEP \textbf{0807} (2008), 052,
  \texttt{arXiv:0806.3039} [hep-th].
%%CITATION = ARXIV:0806.3039;%%

\bibitem{Acharya:2008zi}
B.~S. Acharya, K.~Bobkov, G.~L. Kane, J.~Shao, and P.~Kumar, Phys.Rev.
  \textbf{D78} (2008), 065038, \texttt{arXiv:0801.0478} [hep-ph].
%%CITATION = ARXIV:0801.0478;%%

\bibitem{Bouchard:2005ag}
V.~Bouchard and R.~Donagi, Phys.Lett. \textbf{B633} (2006), 783,
  \texttt{arXiv:hep-th/0512149} [hep-th].
%%CITATION = HEP-TH/0512149;%%

\bibitem{Anderson:2011ns}
L.~B. Anderson, J.~Gray, A.~Lukas, and E.~Palti, Phys.Rev. \textbf{D84} (2011),
  106005, \texttt{arXiv:1106.4804} [hep-th].
%%CITATION = ARXIV:1106.4804;%%

\bibitem{Blaszczyk:2010db}
M.~Blaszczyk, S.~Nibbelink~Groot, F.~Ruehle, M.~Trapletti, and
  P.~K.~S.~Vaudrevange, JHEP \textbf{1009} (2010), 065,
  \texttt{arXiv:1007.0203} [hep-th].
%%CITATION = ARXIV:1007.0203;%%

\bibitem{Blaszczyk:2011ig}
M.~Blaszczyk, N.~G.~C. Bizet, H.~P. Nilles, and F.~Ruehle, JHEP \textbf{1110}
  (2011), 117, \texttt{arXiv:1108.0667} [hep-th].

\bibitem{Blaszczyk:2011hs}
M.~Blaszczyk, S.~Groot~Nibbelink, and F.~Ruehle, JHEP \textbf{1205} (2012),
  053, \texttt{arXiv:1111.5852} [hep-th].
%%CITATION = ARXIV:1111.5852;%%

\bibitem{Buchmuller:2012mu}
W.~Buchm{\"u}ller, J.~Louis, J.~Schmidt, and R.~Valandro, JHEP \textbf{1210}
  (2012), 114, \texttt{arXiv:1208.0704} [hep-th].
%%CITATION = ARXIV:1208.0704;%%

\bibitem{Buchmuller:2005jr}
W.~Buchm{\"u}ller, K.~Hamaguchi, O.~Lebedev, and M.~Ratz, Phys. Rev. Lett.
  \textbf{96} (2006), 121602, \texttt{hep-ph/0511035}.
%%CITATION = HEP-PH 0511035;%%

\bibitem{Buchmuller:2006ik}
W.~Buchm{\"u}ller, K.~Hamaguchi, O.~Lebedev, and M.~Ratz, Nucl. Phys.
  \textbf{B785} (2007), 149, \texttt{hep-th/0606187}.
%%CITATION = HEP-TH 0606187;%%

\bibitem{Lebedev:2006kn}
O.~Lebedev, H.~P. Nilles, S.~Raby, S.~Ramos-S{\'a}nchez, M.~Ratz, P.~K.~S.
  Vaudrevange, and A.~Wingerter, Phys. Lett. \textbf{B645} (2007), 88,
  \texttt{hep-th/0611095}.
%%CITATION = HEP-TH 0611095;%%

\bibitem{Lebedev:2008un}
O.~Lebedev, H.~P. Nilles, S.~Ramos-S\'{a}nchez, M.~Ratz, and P.~K.~S.
  Vaudrevange, Phys. Lett. \textbf{B668} (2008), 331, \texttt{arXiv:0807.4384}
  [hep-th].
%%CITATION = 0807.4384;%%

\bibitem{Lebedev:2009ag}
O.~Lebedev and S.~Ramos-S{\'a}nchez, Phys.Lett. \textbf{B684} (2010), 48,
  \texttt{arXiv:0912.0477} [hep-ph].
%%CITATION = ARXIV:0912.0477;%%

\bibitem{Antoniadis:1994hg}
I.~Antoniadis, E.~Gava, K.~S. Narain, and T.~R. Taylor, Nucl. Phys.
  \textbf{B432} (1994), 187, \texttt{hep-th/9405024}.
%%CITATION = HEP-TH 9405024;%%

\bibitem{Kappl:2008ie}
R.~Kappl, H.~P. Nilles, S.~Ramos-S{\'a}nchez, M.~Ratz, K.~Schmidt-Hoberg, and
  P.~K.~S. Vaudrevange, Phys. Rev. Lett. \textbf{102} (2009), 121602,
  \texttt{arXiv:0812.2120} [hep-th].
%%CITATION = 0812.2120;%%

\bibitem{Kobayashi:2006wq}
T.~Kobayashi, H.~P. Nilles, F.~Pl{\"o}ger, S.~Raby, and M.~Ratz, Nucl.Phys.
  \textbf{B768} (2007), 135, \texttt{arXiv:hep-ph/0611020} [hep-ph].
%%CITATION = HEP-PH/0611020;%%

\bibitem{Lee:2010gv}
H.~M. Lee, S.~Raby, M.~Ratz, G.~G. Ross, R.~Schieren, K.~Schmidt-Hoberg, and
  P.~K.~S. Vaudrevange, Phys.Lett. \textbf{B694} (2011), 491,
  \texttt{arXiv:1009.0905} [hep-ph].

\bibitem{Forste:2010pf}
S.~F{\"o}rste, H.~P. Nilles, S.~Ramos-S{\'a}nchez, and P.~K.~S.~Vaudrevange,
  Phys.Lett. \textbf{B693} (2010), 386, \texttt{arXiv:1007.3915} [hep-ph].
%%CITATION = ARXIV:1007.3915;%%

\bibitem{Kobayashi:1991rp}
T.~Kobayashi and N.~Ohtsubo, Int. J. Mod. Phys. \textbf{A9} (1994), 87.
%%CITATION = IMPAE,A9,87;%%

\bibitem{Bailin:1999nk}
D.~Bailin and A.~Love, Phys.Rept. \textbf{315} (1999), 285.
%%CITATION = PRPLC,315,285;%%

\bibitem{Blumenhagen:2006ab}
R.~Blumenhagen and E.~Plauschinn, JHEP \textbf{0608} (2006), 031,
  \texttt{arXiv:hep-th/0604033} [hep-th].
%%CITATION = HEP-TH/0604033;%%

\bibitem{Donagi:2008xy}
R.~Donagi and K.~Wendland, J.Geom.Phys. \textbf{59} (2009), 942,
  \texttt{arXiv:0809.0330} [hep-th].
%%CITATION = ARXIV:0809.0330;%%

\bibitem{Donagi:2004ht}
R.~Donagi and A.~E. Faraggi, Nucl.Phys. \textbf{B694} (2004), 187,
  \texttt{arXiv:hep-th/0403272} [hep-th].
%%CITATION = HEP-TH/0403272;%%

\bibitem{Forste:2006wq}
S.~F{\"o}rste, T.~Kobayashi, H.~Ohki, and K.-j. Takahashi, JHEP \textbf{0703}
  (2007), 011, \texttt{arXiv:hep-th/0612044} [hep-th].
%%CITATION = HEP-TH/0612044;%%

\bibitem{Beye:2013moa}
F.~Beye, T.~Kobayashi, and S.~Kuwakino, \texttt{arXiv:1304.5621} [hep-th].
%%CITATION = ARXIV:1304.5621;%%

\bibitem{Fischer:2012qj}
M.~Fischer, M.~Ratz, J.~Torrado, and P.~K.~S.~Vaudrevange, JHEP \textbf{1301}
  (2013), 084, \texttt{arXiv:1209.3906} [hep-th].
%%CITATION = ARXIV:1209.3906;%%

\bibitem{Kakushadze:1996hj}
Z.~Kakushadze, G.~Shiu, and S.~H. Tye, Phys.Rev. \textbf{D54} (1996), 7545,
  \texttt{arXiv:hep-th/9607137} [hep-th].
%%CITATION = HEP-TH/9607137;%%

\bibitem{Konopka:2012gy}
S.~J. Konopka, \texttt{arXiv:1210.5040} [hep-th].
%%CITATION = ARXIV:1210.5040;%%

\bibitem{Ross:2004mi}
G.~Ross, \texttt{arXiv:hep-ph/0411057} [hep-ph].
%%CITATION = HEP-PH/0411057;%%

\bibitem{Hebecker:2004ce}
A.~Hebecker and M.~Trapletti, Nucl.Phys. \textbf{B713} (2005), 173,
  \texttt{arXiv:hep-th/0411131} [hep-th].
%%CITATION = HEP-TH/0411131;%%

\bibitem{Anandakrishnan:2012ii}
A.~Anandakrishnan and S.~Raby, Nucl.Phys. \textbf{B868} (2013), 627,
  \texttt{arXiv:1205.1228} [hep-ph].
%%CITATION = ARXIV:1205.1228;%%

\bibitem{Kobayashi:2004ya}
T.~Kobayashi, S.~Raby, and R.-J. Zhang, Nucl. Phys. \textbf{B704} (2005), 3,
  \texttt{hep-ph/0409098}.
%%CITATION = HEP-PH 0409098;%%

\bibitem{Kim:2007mt}
J.~E. Kim, J.-H. Kim, and B.~Kyae, JHEP \textbf{06} (2007), 034,
  \texttt{hep-ph/0702278}.
%%CITATION = HEP-PH/0702278;%%

\bibitem{Blaszczyk:2009in}
M.~Blaszczyk, S.~G. Nibbelink, M.~Ratz, F.~Ruehle, M.~Trapletti, et~al.,
  Phys.Lett. \textbf{B683} (2010), 340, \texttt{arXiv:0911.4905} [hep-th].

\bibitem{Pena:2012ki}
D.~K.~M. Pena, H.~P. Nilles, and P.-K. Oehlmann, JHEP \textbf{1212} (2012),
  024, \texttt{arXiv:1209.6041} [hep-th].
%%CITATION = ARXIV:1209.6041;%%

\bibitem{Ibanez:1987pj}
L.~E. Ib{\'a}{\~n}ez, J.~Mas, H.-P. Nilles, and F.~Quevedo, Nucl.Phys.
  \textbf{B301} (1988), 157.
%%CITATION = NUPHA,B301,157;%%

\bibitem{Casas:1990yt}
J.~Casas, A.~de~la Macorra, M.~Mondrag{\'o}n, and C.~Mu{\~n}oz, Phys.Lett.
  \textbf{B247} (1990), 50.
%%CITATION = PHLTA,B247,50;%%

\bibitem{Brown:1978}
Brown, Buelow, Neubueser, Wondratchek, and Zassenhaus.

\bibitem{Plesken:1998}
J.~Opgenorth, W.~Plesken, and T.~Schulz, Acta Cryst. Sect. A \textbf{54}
  (1998).

\bibitem{GAP4}
The GAP~Group, \textit{{GAP -- Groups, Algorithms, and Programming, Version
  4.6.3}}, 2013.

\bibitem{Ellis:1983sf}
J.~R. Ellis, A.~Lahanas, D.~V. Nanopoulos, and K.~Tamvakis, Phys.Lett.
  \textbf{B134} (1984), 429.
%%CITATION = PHLTA,B134,429;%%

\bibitem{Cvetic:1988yw}
M.~Cveti{\v{c}}, J.~Louis, and B.~A. Ovrut, Phys. Lett. \textbf{B206} (1988),
  227.
%%CITATION = PHLTA,B206,227;%%

\bibitem{Brignole:1997dp}
A.~Brignole, L.~E. Ibanez, and C.~Munoz, \texttt{arXiv:hep-ph/9707209}
  [hep-ph].
%%CITATION = HEP-PH/9707209;%%

\bibitem{Covi:2008ea}
L.~Covi, M.~Gomez-Reino, C.~Gross, J.~Louis, G.~A. Palma, et~al., JHEP
  \textbf{0806} (2008), 057, \texttt{arXiv:0804.1073} [hep-th].
%%CITATION = ARXIV:0804.1073;%%

\bibitem{Ibanez:1992hc}
L.~E. Ib{\'a}{\~n}ez and D.~L{\"u}st, Nucl.Phys. \textbf{B382} (1992), 305,
  \texttt{arXiv:hep-th/9202046} [hep-th].
%%CITATION = HEP-TH/9202046;%%

\bibitem{Dine:1986zy}
M.~Dine, N.~Seiberg, X.~Wen, and E.~Witten, Nucl.Phys. \textbf{B278} (1986),
  769.
%%CITATION = NUPHA,B278,769;%%

\bibitem{Dixon:1989fj}
L.~J. Dixon, V.~Kaplunovsky, and J.~Louis, Nucl.Phys. \textbf{B329} (1990), 27.
%%CITATION = NUPHA,B329,27;%%

\bibitem{Nibbelink:2012de}
S.~Groot~Nibbelink and P.~K.~S.~Vaudrevange, JHEP \textbf{1303} (2013), 142,
  \texttt{arXiv:1212.4033} [hep-th].
%%CITATION = ARXIV:1212.4033;%%

\bibitem{Nilles:2011aj}
H.~P. Nilles, S.~Ramos-S{\'a}nchez, P.~K.~S.~Vaudrevange, and A.~Wingerter,
  Comput.Phys.Commun. \textbf{183} (2012), 1363, \texttt{arXiv:1110.5229}
  [hep-th], web page http://projects.hepforge.org/orbifolder/.
%%CITATION = ARXIV:1110.5229;%%

\bibitem{Ibanez:1986tp}
L.~E. Ib{\'a}{\~n}ez, H.~P. Nilles, and F.~Quevedo, Phys.Lett. \textbf{B187}
  (1987), 25.
%%CITATION = PHLTA,B187,25;%%

\bibitem{Brown:2002}
R.~{Brown} and P.~J. {Higgins}, ArXiv Mathematics e-prints (2002),
  \texttt{arXiv:math/0212271}.

\bibitem{Anderson:2009mh}
L.~B. Anderson, J.~Gray, Y.-H. He, and A.~Lukas, JHEP \textbf{1002} (2010),
  054, \texttt{arXiv:0911.1569} [hep-th].

\bibitem{Braun:2010vc}
V.~Braun, JHEP \textbf{1104} (2011), 005, \texttt{arXiv:1003.3235} [hep-th].
%%CITATION = ARXIV:1003.3235;%%

\bibitem{Witten:1996mz}
E.~Witten, Nucl.Phys. \textbf{B471} (1996), 135, \texttt{arXiv:hep-th/9602070}
  [hep-th].
%%CITATION = HEP-TH/9602070;%%

\bibitem{Abe:2009uz}
H.~Abe, K.-S. Choi, T.~Kobayashi, and H.~Ohki, Phys.Rev. \textbf{D80} (2009),
  126006, \texttt{arXiv:0907.5274} [hep-th].
%%CITATION = ARXIV:0907.5274;%%

\bibitem{KashaniPoor:2013en}
A.-K. Kashani-Poor, R.~Minasian, and H.~Triendl, JHEP \textbf{1304} (2013),
  058, \texttt{arXiv:1301.5031} [hep-th].
%%CITATION = ARXIV:1301.5031;%%

\bibitem{Ishimori:2012zz}
H.~Ishimori, T.~Kobayashi, H.~Ohki, H.~Okada, Y.~Shimizu, et~al., Lect.Notes
  Phys. \textbf{858} (2012), pp.1.
%%CITATION = LNPHA,858,pp.1;%%

\bibitem{Luhn:2007sy}
C.~Luhn, S.~Nasri, and P.~Ramond, Phys.Lett. \textbf{B652} (2007), 27,
  \texttt{arXiv:0706.2341} [hep-ph].
%%CITATION = ARXIV:0706.2341;%%

\bibitem{Luhn:2012bc}
C.~Luhn, K.~M. Parattu, and A.~Wingerter, JHEP \textbf{1212} (2012), 096,
  \texttt{arXiv:1210.1197} [hep-ph].
%%CITATION = ARXIV:1210.1197;%%

\bibitem{Blumenhagen:1999ev}
R.~Blumenhagen, L.~Gorlich, and B.~Kors, JHEP \textbf{0001} (2000), 040,
  \texttt{arXiv:hep-th/9912204} [hep-th].
%%CITATION = HEP-TH/9912204;%%

\bibitem{Forste:2000hx}
S.~F{\"o}rste, G.~Honecker, and R.~Schreyer, Nucl.Phys. \textbf{B593} (2001),
  127, \texttt{arXiv:hep-th/0008250} [hep-th].
%%CITATION = HEP-TH/0008250;%%

\bibitem{Erler:1993zy}
J.~Erler, J.Math.Phys. \textbf{35} (1994), 1819, \texttt{arXiv:hep-th/9304104}
  [hep-th].
%%CITATION = HEP-TH/9304104;%%

\bibitem{Honecker:2006qz}
G.~Honecker and M.~Trapletti, JHEP \textbf{0701} (2007), 051,
  \texttt{arXiv:hep-th/0612030} [hep-th].
%%CITATION = HEP-TH/0612030;%%

\end{thebibliography}
\bibliographystyle{NewArXiv.bst}
\end{document}